\documentclass[useAMS,usenatbib]{mn2e}
\usepackage{psfig}
\usepackage{epsfig}
\usepackage{graphicx}
\usepackage{multirow}
\usepackage{subfigure}

\title[SSP models with BSs]
{Simple stellar population models including blue stragglers}
\author[Y. Xin et al.]{Yu Xin,$^{1}$\thanks{YX is supported by the Alexander von Humboldt Foundation. 
E-mail: yxin@astro.uni-bonn.de} Licai Deng,$^{2}$ Richard de~Grijs$^{3}$ 
and Pavel Kroupa$^{1}$ \\
$^{1}$Argelander-Institut f\"{u}r Astronomie (Sternwarte),
Universit\"{a}t Bonn, Auf dem H\"{u}gel 71, D-53121 Bonn, Germany\\
$^{2}$Key laboratory for optical astronomy, National Astronomical Observatories, Chinese Academy of
Sciences, Beijing 100012, China\\
$^{3}$Kavli Institute for Astronomy and Astrophysics, 
Peking University, Beijing 100871, China}

\begin{document}

\date{Accepted ---. Received ---; in original form ---}


\maketitle


\begin{abstract}
Observations show that nearly all star clusters and stellar
populations contain blue straggler stars (BSs). 
BSs in a cluster can significantly enhance the integrated
spectrum of the host population, preferentially at short wavelengths,
and render it much bluer in photometric colours. Current theoretical
simple stellar population (SSP) models constructed within the
traditional framework of single and binary stellar evolution cannot
fully account for the impact of these objects on the integrated
spectral properties of stellar populations. Using conventional SSP
models without taking into account BS contributions may significantly
underestimate a cluster's age and/or metallicity, simply because one
has to balance the observed bluer colours (or a bluer spectrum) with a
younger age and/or a lower metallicity. Therefore, inclusion of BS
contributions in SSP models is an important and necessary improvement
for population synthesis and its applications. Here, we present a new
set of SSP models, which include BS contributions based on our
analysis of individual star clusters. The models cover the wavelength
range from 91~{\AA} to 160~$\mu$m, ages from 0.1 to 20 Gyr and
metallicities $Z=0.0004, 0.004, 0.008, 0.02$ (solar) and 0.05. We use
the observed integrated spectra of several Magellanic Cloud star
clusters to cross-check and validate our models. The results show that
the age predictions from our models are closer to those from isochrone
fitting in the clusters' colour-magnitude diagrams compared to age
predictions based on standard SSP models.
\end{abstract}


\begin{table*}
\caption{Fundamental ingredients of the BS-SSP models.}
\label{table1}
\begin{tabular}{lll}
\hline
\hline
Name & Property & Source\\
\hline 
Padova1994 isochrones & $Z$=0.0001--0.05~~Age=4Myr--20Gyr & Bertelli et al. (1994)\\
BaSeL spectral library  & 91{\AA}--160$\mu$m & Lejeune et al. (1997)\\
                                        & median resolution $\lambda/\Delta\lambda\approx300$ &\\
Initial mass function & $\xi(\log~m)\propto m^{-1.35}$ & Salpeter IMF (Salpeter 1955)  \\  
                                     & table~1 in Chabrier (2003)        & Canonical IMF (Kroupa 2001, 2002; Chabrier 2003)\\
\hline
\end{tabular}
\end{table*}

\section{Introduction}

Evolutionary population synthesis (EPS) has been widely used as a
powerful tool to study the stellar contents of galaxies. In essence,
EPS compares the observed, integrated spectrum of a galaxy with a
combination of spectra of simple stellar populations (SSPs;
single-age, single-metallicity populations) of different ages and
metallicities to infer the galaxy's star-formation history. Over the
past two decades, much work has been done to improve the accuracy of
EPS and SSP models in various contexts (e.g., Bica \& Alloin 1986;
Fritze-v. Alvensleben \& Gerhard 1994; Worthey 1994; Leitherer et
al. 1999; Vazdekis 1999; Bruzual \& Charlot 2003, hereafter BC03;
Thomas et al. 2003). Unfortunately, population synthesis models still
suffer from a number of limitations. One is our poor understanding of
some advanced single-star evolutionary phases, such as of supergiants
and asymptotic-giant-branch (AGB) stars (Yi 2003), while a second is an
absence in the models of the results of stellar interactions, such as
the so-called `stragglers' formed through mass transfer in binaries or
stellar collisions. Such stars are usually very bright and can
strongly affect the integrated-light properties of the entire
system. The potential uncertainties inherent to EPS caused by ignoring
these components could be much larger than those still remaining and
due to the variety of input physics among different models.

In this paper, we focus on the second limitation to the standard SSP
models. We present a new set of SSP models which include contributions
from blue straggler stars (BSs).

BSs are common and easily identified in colour-magnitude diagrams
(CMDs) of star clusters. They are members of the host cluster and
located above and blueward of the cluster's main-sequence (MS)
turnoff. The standard theory of single-star evolution cannot explain
the presence of BSs in SSP CMDs, and thus the standard SSP models do
not include contributions of BSs. All currently accepted scenarios of
BS formation are related to stellar interactions. Coalescence in
primordial binaries can launch BSs to positions up to 2.5 magnitudes
brighter than the MS turnoff (McCrea 1964). Mergers of binary-binary
systems can produce possible BSs with masses four times those of stars
at the MS turnoff (Leonard \& Linnell 1992). Stellar collisions
are also an important formation channel of BS formation in globular
clusters (GCs) and open clusters (OCs; Glebbeek et al. 2008). Given
the high luminosities and common presence of BSs in stellar systems
(e.g., Ahumada \& Lapasset 1995 for OCs; Piotto et al. 2002 for GCs;
Mapelli et al. 2009 for dwarf galaxies), we believe that we must
consider the effects of BSs in studies of stellar populations using
population synthesis applied to unresolved observations. The key issue
is how to accurately include BS contributions in SSP models.

Studies show that no single mechanism can account for the entire BS
population observed in any one star cluster (Stryker 1993). This means
that it is not easy to theoretically measure the respective
contribution of BSs in SSPs of different ages and
metallicities. Therefore, building up BS population characteristics
empirically from the statistics of a large sample of star clusters
could be more practical and reliable than relying on incomplete
theoretical approaches. This way, the behaviour of BSs (in terms of
their specific frequency and relative distribution with respect to the
MS turnoff in CMDs) can be modelled. OCs in the Galaxy have a number
of advantages for use as a working sample, for instance, (i) they
are good observational templates of ideal SSPs in the real world, and
all BSs in a given OC belong to a single population; (ii) many OCs
have multi-epoch proper-motion and/or radial-velocity data, so that
their cluster membership probabilities can be measured accurately.

Preliminary modelling of BS effects has been done on the basis of
individual clusters in our previous papers (Deng et al. 1999; Xin \&
Deng 2005; Xin et al. 2007, 2008), where we (i) introduced the method
used for calculating the realistic, integrated spectrum of a star
cluster including the contribution of BSs; (ii) analysed the
modifications to the integrated spectra and broad-band colours caused
by BSs; and (iii) estimated the possible uncertainties in the
conventional SSP models, showing that the ages of star clusters can be
underestimated by up to 50\% if BS contributions are not considered.

In this paper, we present a set of BS-SSP models based on a
statistical study of Galactic OCs. A working sample including 100
Galactic OCs is used to reduce stochastic effects in collecting the
required statistical relations.  We also use the standard deviation
($\sigma$) to keep the statistics in a relatively small dispersion
region. Our models cover the wavelength range from 91 {\AA} to 160
$\mu$m, ages from 0.1 to 20 Gyr and metallicities $Z=0.0004, 0.004,
0.008, 0.02$ (solar metallicity) and 0.05.

This paper is organised as follows. In Section 2, we present the
statistical results of the properties of BS populations based on 100
Galactic OCs. Here, we also describe the construction procedure of our
models. In Section 3, we discuss our model results and compare them
with those of BC03. In Section 4, we test our models by comparison
with the broad-band colours of OCs. Our model colours are in good
agreement with observed OC colours (which include BS contributions).
In Section 5, we test our models using observed integrated
spectra of star clusters. Compared to BC03, our models yield more
accurate age predictions at a range of wavelengths, including in the
optical regime. Finally, a summary and further brief discussion is
contained in Section 6.

\section{BS-SSP models}

The fundamental ingredients of our models are listed in
Table~\ref{table1}. For convenience, we use the widely adopted BC03
models as reference. Modifications owing to BSs are calculated as
increments to the BC03 SSPs of the same age and metallicity. We
adopted the Padova1994 isochrones (Bertelli et al. 1994) and the BaSeL
spectral library (Lejeune et al. 1997) because they homogeneously
cover the widest ranges of age and metallicity, and the longest
wavelength range. The Padova2000 isochrones (Girardi et al. 2000) are
based on a more recent equation of state and low-temperature opacities
compared to Padova1994. However, we decided against adopting them for
our model construction, because BC03 do not recommend to use their
SSPs based on the Padova2000 isochrones.  They state that their models
based on the Padova2000 isochrones ``tend to produce worse agreement
with observed galaxy colours'' (BC03, their
footnote~6). High-resolution observational spectral libraries (e.g.,
Pickles et al. 1998; Le~Borgne et al. 2003) suffer from problems
related to limited parameter coverage. Instead of combining spectra
from different libraries to enlarge our parameter coverage, we decided
to use only the theoretical library.

\begin{figure}
 \includegraphics[width=8cm]{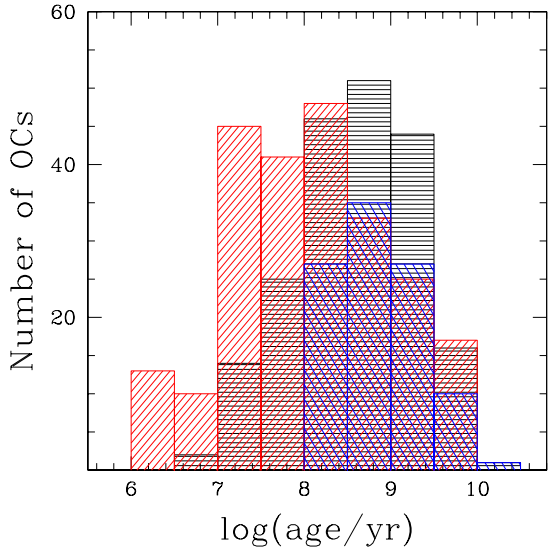}
\caption{Number distributions of Galactic OCs containing BSs as a
  function of age. Horizontal shading represents the statistics from
  the Ahumada \& Lapasset (2007) catalogue. The hatching from the
  bottom left to the top right represents the statistics from the
  Ahumada \& Lapasset (1995) catalogue, while that from the top left
  to the bottom right shows the number distribution of our working
  sample (100 Galactic OCs).}
\label{fig1}
\end{figure}

\begin{figure}
\includegraphics[width=4.4cm]{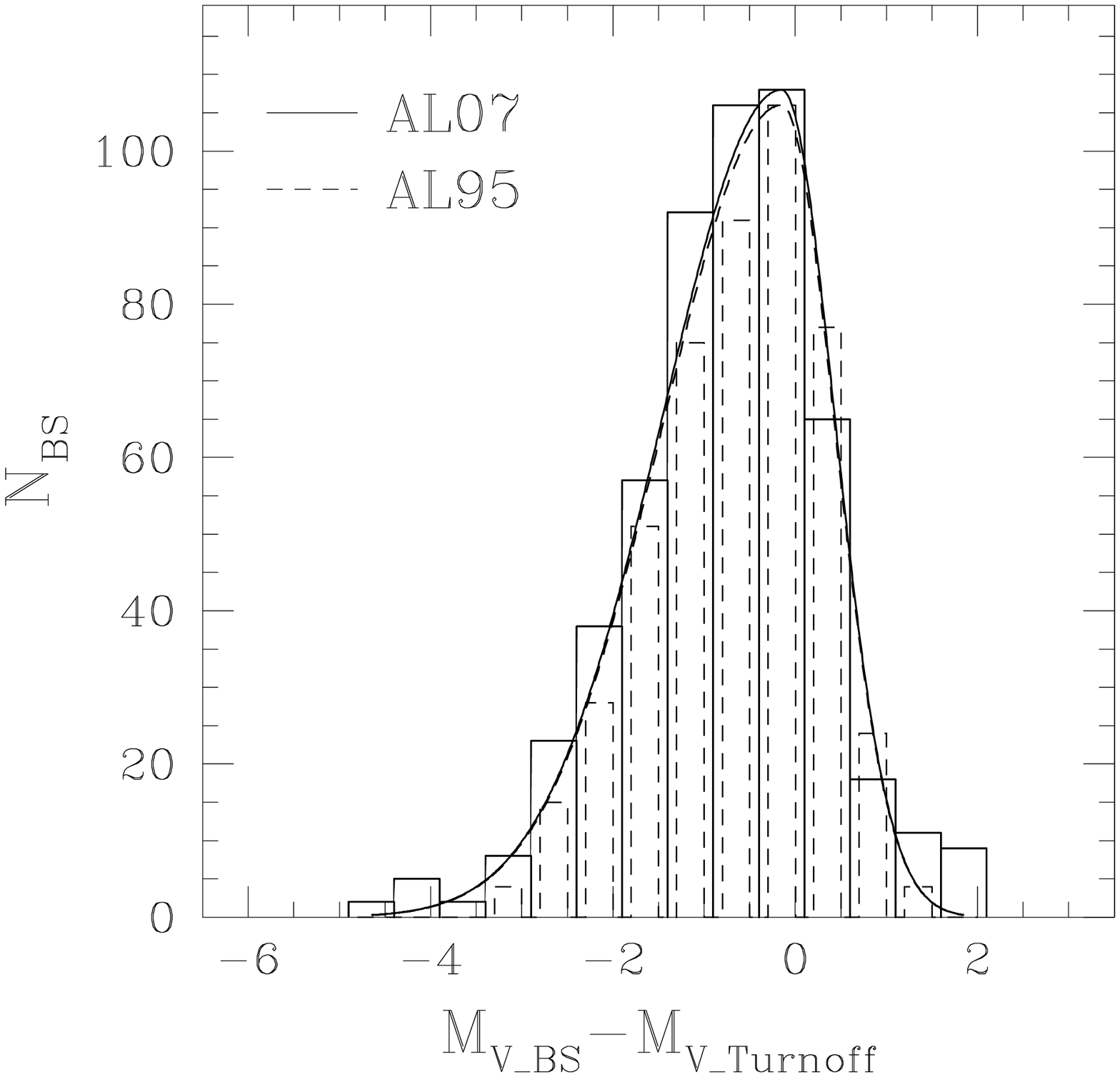}\includegraphics[width=4.4cm]{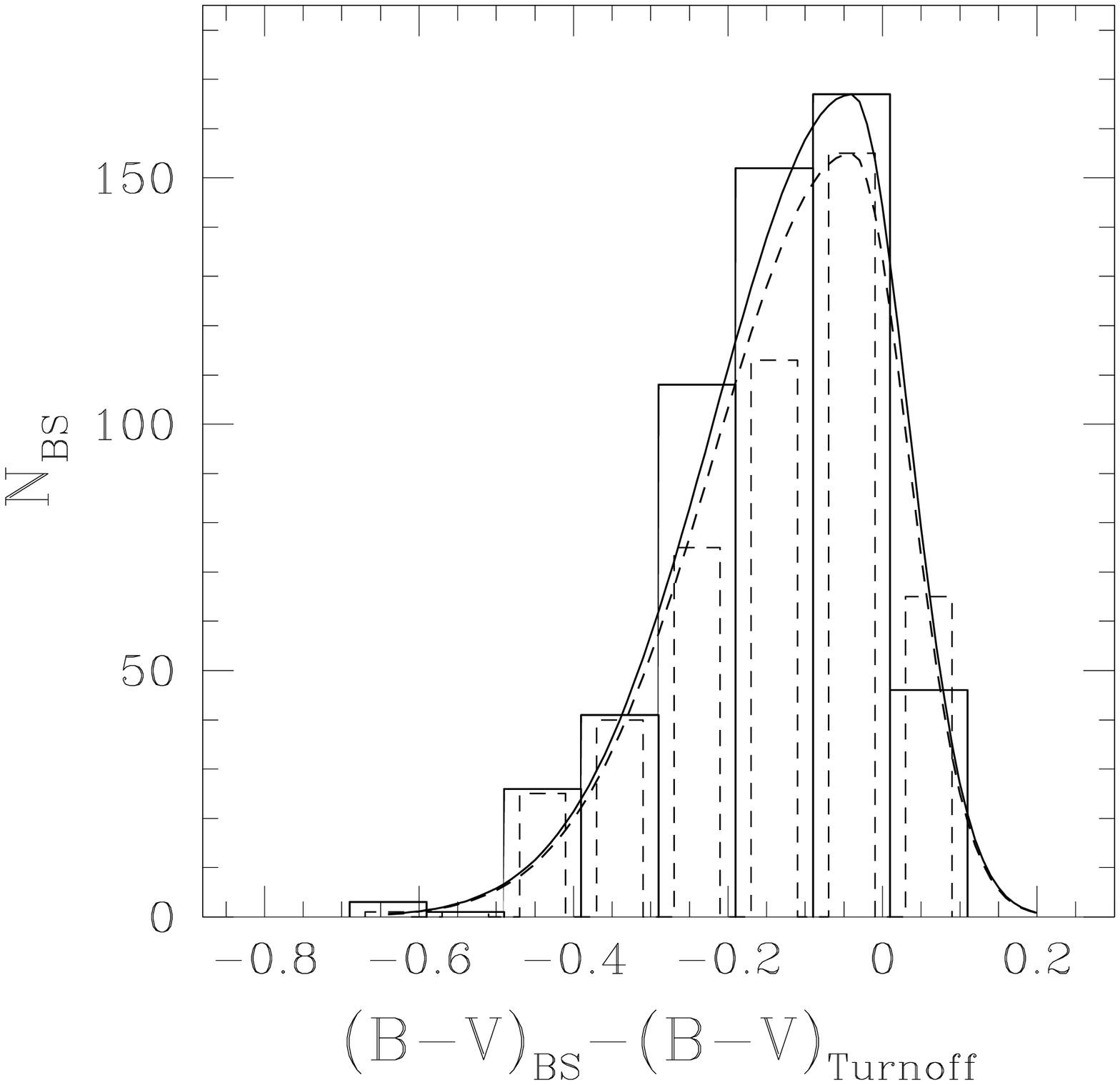}
\caption{BS distribution functions in the CMD versus $M_V$ (left
  panel) and $(B-V)$ (right panel). The solid histograms represent the
  statistics from the AL07 catalogue, and the solid curves are the
  corresponding Gaussian approximations to either side of the
  peak. The dashed histograms and dashed curves are the results from
  the AL95 catalogue. The statistics are processed based on the
  photometric data of BSs in 33 OCs of age$\geq1.0$ Gyr and common to
  all of AL95, AL07 and our working sample.}
\label{fig2}
\end{figure}

For consistency with BC03, we adopted the Salpeter and Chabrier
stellar initial mass functions (IMFs; Salpeter 1955; Chabrier
2003). As clearly shown by Dabringhausen et al. (2008, their fig. 8),
the Chabrier IMF is almost indistinguishable from the Kroupa IMF
(Kroupa 2001, 2002) when normalised as $\int_{0.1}^{100}\xi(m){\rm
  d}m=1{\rm M}_{\odot}$, which is exactly how both BC03 and we
ourselves normalise the SSP models. Using such a normalisation, the
slight differences between the two IMFs cannot cause any effective
modifications as regards the BS contribution to SSPs. Therefore, we
refer to the IMF (in addition to the Salpeter IMF) as the `Canonical
IMF' throughout this work. It can be conveniently described by a
two-part power law, $\xi(m) \propto m^{-\alpha_i}$, with
$\alpha_1=1.3$ for the stellar-mass range $0.08 \le m/{\rm M}_\odot <
0.5$ and $\alpha_2=2.3$ for $m \ge 0.5 {\rm M}_{\odot}$ (Kroupa 2001),
or in terms of a power law plus a lognormal form as presented in
table~1 of Chabrier (2003).

\subsection{BS statistics}

Basically, two properties of BS populations are relevant to the
modelling of BS behaviour in SSPs, i.e., the number of BSs ($N_{\rm
  BS}$) in the SSP and their distribution in the SSP's CMD. In this
paper, both properties are obtained empirically from the observed OC
CMDs. A working sample including 100 Galactic OCs from the catalogue
of Ahumada \& Lapasset (1995, hereafter AL95) is adopted to secure the
reliability of the statistical results.

Fig.~\ref{fig1} shows the differences among the number distributions
of OCs containing BSs versus age for the Ahumada \& Lapasset (2007,
hereafter AL07) catalogue (horizontal hatching), the AL95 catalogue
(slanted hatching from bottom left to top right) and our sample
(slanted hatching from top left to bottom right). AL95 and AL07
published the most complete catalogues to date in terms of photometric
data of BSs in Galactic OCs. They include almost all OCs containing
BSs in the solar neighbourhood. In particular, they also include
the so-called `yellow stragglers' (e.g., Portegies~Zwart et al. 1997;
Deng et al. 1999, their fig. 1) in their BS catalogues. Yellow
stragglers are located between the MS turnoff and the giant branch in
CMDs (i.e., they are redder than BSs), therefore adding extra light
preferentially in the optical spectral regions of a cluster.

Fig.~\ref{fig1} shows that, compared to AL95, AL07 dramatically
reduced the number of young OCs containing BSs, mainly because of the
difficulty of identifying BSs in the CMD of a star cluster that does
not exhibit at least a fully developed red-giant-branch (RGB) phase
(particularly when membership-probability information is
lacking). Moreover, most of the BSs in young OCs (i.e.,
log(age/yr)$<$8.0 in this paper) from both catalogues are located
close to the MS turnoff point in the CMD, which means that it is hard
to distinguish BSs from MS stars. Such BSs cannot effectively modify
the spectral intensity of a cluster. Therefore, we start the selection
of our working sample based on OCs with log(age/yr)$\geq$8.0.

Meanwhile, to construct a spectrum of the BS population for a given
OC, we need to obtain the physical parameters of the BSs in the CMD of
the OC (see also Section~2.2). To do this, we need the age,
metallicity, colour excess and distance modulus for each OC, which are
not all included in either AL95 or AL07. Therefore, we decided to keep
only the photometric data of BSs from AL95, to ensure a homogeneous
selection of the BS sample. We collected the remaining OC parameters
from the recent literature (see Xin et al. 2007, their table 1). In
practice, OCs with log(age/yr)$\geq$8.0 from AL95 are included as
sample clusters if reliable parameters can be found, in the sense that
most of the BSs are located at reasonable positions in the CMD with
respect to the Padova1994 isochrone for the OC's age and metallicity.

There are 33 OCs with age$\ge$1.0 Gyr in the combined catalogue
comprised of AL95, AL07 and our working sample. Older OCs have better
statistics as regards their BS populations. Using the photometric data
of the BSs in these 33 OCs as an example, we show the BS distribution
functions versus the MS turnoff point in Fig.~\ref{fig2}, as a
function of $M_V$ (left panel) and $(B-V)$ (right panel). We use this
figure to detect differences in the BS distributions in the CMD
between the AL95 and AL07 statistics. The solid histograms represent
the statistics from the AL07 catalogue, and the solid curves are the
corresponding Gaussian approximations to either side of the peak. The
dashed histograms and dashed curves are the results from the AL95
catalogue. The same technique is used to construct the BS distribution
functions in the CMD for our model construction (see below). Based on
Fig.~\ref{fig2}, the distribution functions of both $M_V$ and $(B-V)$
do not exhibit any essential differences between the AL95 and the AL07
catalogues. Therefore, we continue on the basis of the results from
our previous work, i.e., the parameters of the sample clusters and the
BS population from AL95.
\begin{figure*}
\begin{center}
\includegraphics[width=15cm]{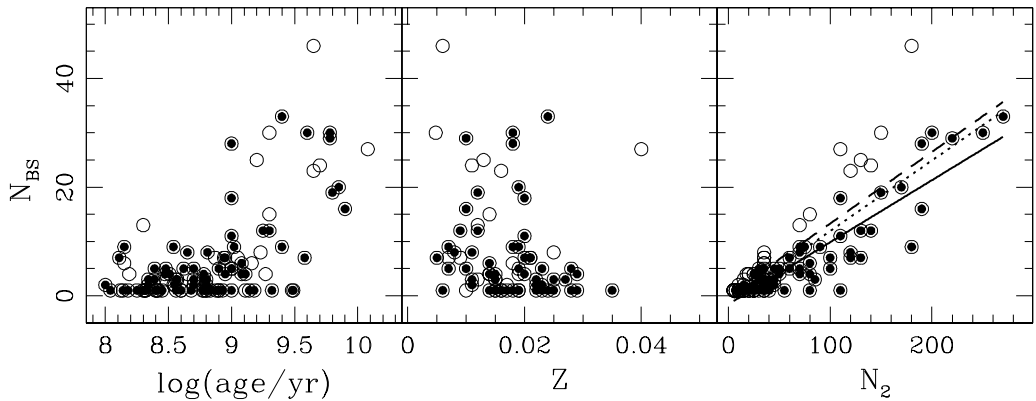}
\caption{Correlations (if any) between $N_{\rm BS}$ and the age,
  metallicity and richness ($N_2$) of Galactic OCs, respectively. The
  open circles show the results for the 100 OCs in our working
  sample. The filled circles mark the OCs with $N_{\rm BS}/N_2$ less
  than the average ratio plus $1\sigma$ (standard deviation). The
  solid line in the right-hand panel is the least-squares fit to the
  filled circles; it is given by Eq.~(\ref{eq1}). For reference, the
  dotted line in the right-hand panel is the least-squares fit to all
  OCs (open circles), and the dashed line is the fit to OCs with
  ages$\geq$1.0 Gyr.}
\label{fig3}
\end{center}
\end{figure*}
Details of the BS properties are presented in
Figs.~\ref{fig3}--\ref{fig5}. Fig.~\ref{fig3} shows $N_{\rm BS}$ as a
function of the age, metallicity and richness of the sample OCs. The
richness of a star cluster is represented by $N_2$, which is the
number of cluster member stars within 2 magnitudes below the cluster's
MS turnoff. The open circles represent the results for the 100
Galactic OCs. Because of the small number of member stars and even
smaller number of BSs in the individual OCs, the results directly
derived from Fig.~\ref{fig3} are very stochastic, and thus we use the
ratio of $N_{\rm BS}$/$N_2$ to reduce the effects of stochasticity. We
use this ratio as definition of the specific frequency of BS
components in SSPs. We calculated the standard deviation ($\sigma$) of
the ratio for the entire sample, i.e., 
$\sigma=\sqrt
{\frac{\sum_{i=1}^N(\frac{N_{\rm BS}}{N_2}_i-\overline{\frac{N_{\rm
          BS}}{N_2}}~)^2}{N\times(N-1)}}$ and $N=100$. We marked the
OCs with $\frac{N_{\rm BS}}{N_2}\le \overline{\frac{N_{\rm BS}}{N_2}}
+ 1\sigma$ with filled circles in Fig.~\ref{fig3}.

\begin{figure}
\includegraphics[width=8cm]{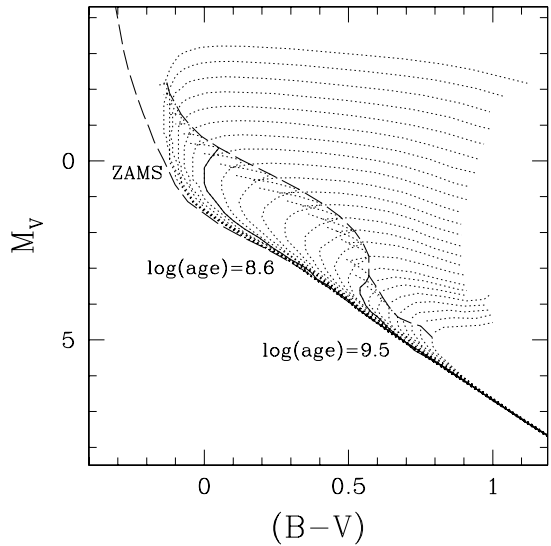}
\caption{Criteria used for our age-bin selection. The dotted lines are
  the isochrones for ages between 0.1 and 20 Gyr, truncated at the
  bottom of the RGB phase. The two dashed lines mark the
  boundaries of the MS stage. The choice of age bin is shown as the
  solid lines.}
\label{fig4}
\end{figure}

\begin{table*}
\caption{Gaussian-profile parameters for BS distribution functions in
  different age bins.}
\label{table2}
\begin{tabular}{cccccccc}
\hline
\hline
Age bin& &  bin size&$\mu$& \multicolumn{2}{c}{$f^-(x)$}&\multicolumn{2}{c}{$f^+(x)$} \\
(yr) & & (mag) & (mag) & $\sigma$  & $x$ range & $\sigma$ & $x$ range\\
\hline
8.0$\leq$log(age/yr)$\leq$8.6&$M_V$&0.45& 0.26 & 0.9454$\pm$0.0126&[$-$3.00, 0.26]&0.4056$\pm$0.0451&[0.26, 1.65]\\
                                      &$(B-V)$ &0.03 & $-$0.01 &0.0563$\pm$ 0.0084&[$-$0.19, $-$0.01] & 0.0418$\pm$0.0027 & [$-$0.01, 0.12]\\
\hline
8.6$<$log(age/yr)$\leq$9.5 &$M_V$&0.40&0.112 & 1.0464$\pm$0.1299&[$-$3.50, 0.112]&0.5470$\pm$0.0532&[0.112, 2.00]\\
                                      &$(B-V)$ &0.075&$-$0.005 &0.1544$\pm$0.0109&[$-$0.530, $-$0.005]&0.0678$\pm$0.0079&[$-$0.005, 0.220]\\
\hline
log(age/yr)$>$9.5&$M_V$&0.55&$-$1.08 & 0.8155$\pm$0.0289 & [$-$3.83, $-$1.08]&0.6711$\pm$0.1931&[$-$1.08, 1.20]\\
                             &$(B-V)$&0.08 & $-$0.06 & 0.1472$\pm$0.0080&[$-$0.55, $-$0.06]&0.0837$\pm$0.0161&[$-$0.06, 0.20]\\
\hline
\hline
\end{tabular}
\end{table*}

The left-hand panel in Fig.~\ref{fig3} shows $N_{\rm BS}$ versus
cluster age on a logarithmic scale. $N_{\rm BS}$ seems largely
insensitive to age until a sudden increase for ages greater than
1.0 Gyr. Based on this figure it is hard to discern any correlation
between the two parameters, and it is also risky to jump to the
conclusion that $N_{\rm BS}$ is not correlated with age. The seemingly
constant $N_{\rm BS}$ for age$<$1.0 Gyr could be caused by the
confusion of defining an accurate MS turnoff point and a BS population
in relatively young star clusters. In our model construction, we have
not adopted any correlation between $N_{\rm BS}$ and age. The only
correlation we used is that between $N_{\rm BS}$ and $N_2$ (shown in
the right-hand panel in Fig.~\ref{fig3}; see below). In fact, for
SSPs, $N_{\rm BS}$ and age are related through the SSP's $N_2$. $N_2$
increases following the IMF slope as the SSP ages, and so does $N_{\rm
  BS}$ through the correlation between $N_{\rm BS}$ and $N_2$.

The middle panel of Fig.~\ref{fig3} shows $N_{\rm BS}$ as a function
of metallicity for our sample OCs with published metallicity
information. No correlation can be established. We previously studied
whether the BS strengths are sensitive to metallicity. We did not find
any obvious correlation between these two parameters either (Xin \&
Deng 2005, their fig. 18).

The right-hand panel of Fig.~\ref{fig3} clearly shows that only
$N_{\rm BS}$ and $N_2$ are correlated. The implication of this
correlation is that $N_{\rm BS}$ is proportional to the richness of an
SSP, but not (at least not obviously) to any other parameter, such as
age or metallicity. The solid line in the right-hand panel is the
least-squares fit to the filled circles. For reference, we present two
more fits for different samples. The dotted line is the least-squares
fit to all OCs (open circles), while the dashed line is the fit to
those OCs that are older than 1.0~Gyr. It is hard to tell which fit is
the most accurate. We choose the solid line for estimating $N_{\rm BS}$ in
an SSP simply to avoid exaggeration of the BS-enhanced intensity in
our SSPs. The correlation can be empirically described as
\begin{equation}
N_{\rm BS}= (0.114\pm0.006) \times N_2 - (1.549\pm0.731). \label{eq1} 
\end{equation}
The uncertainties in the coefficients result from the $1\sigma$
uncertainty in $N_{\rm BS}/N_2$.

\begin{figure}
\includegraphics[width=4.3cm]{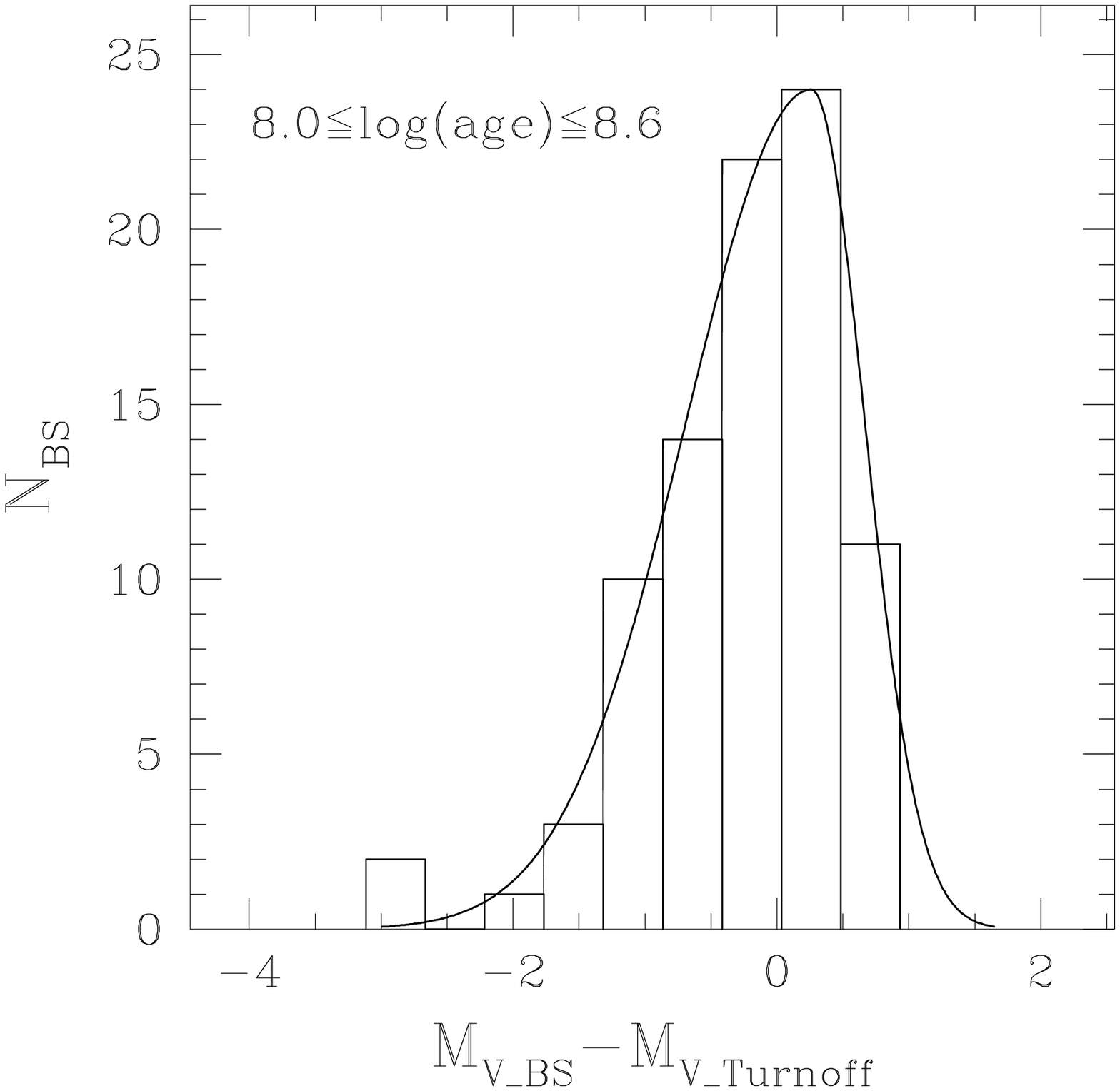}\includegraphics[width=4.3cm]{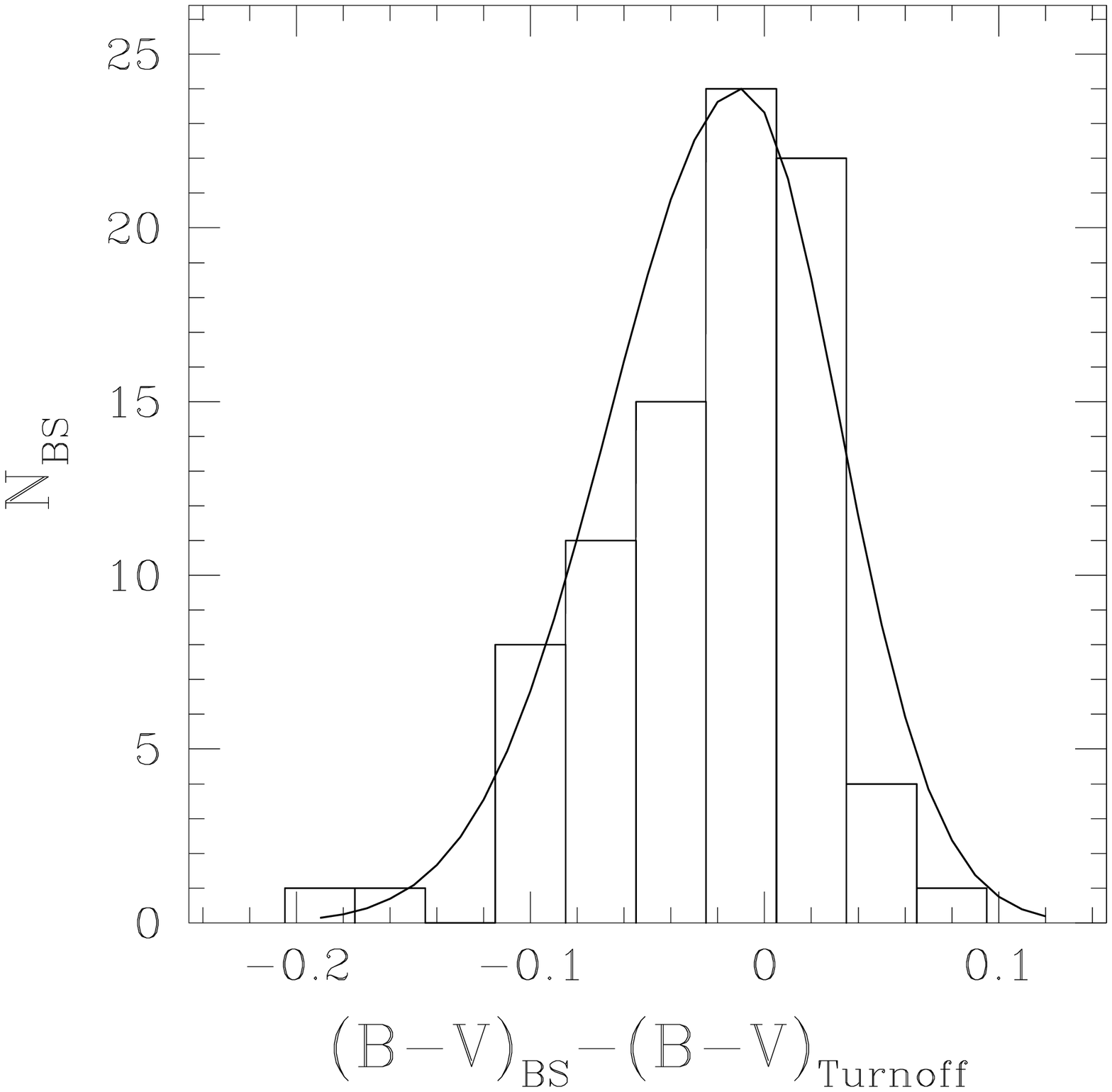}\\
\includegraphics[width=4.3cm]{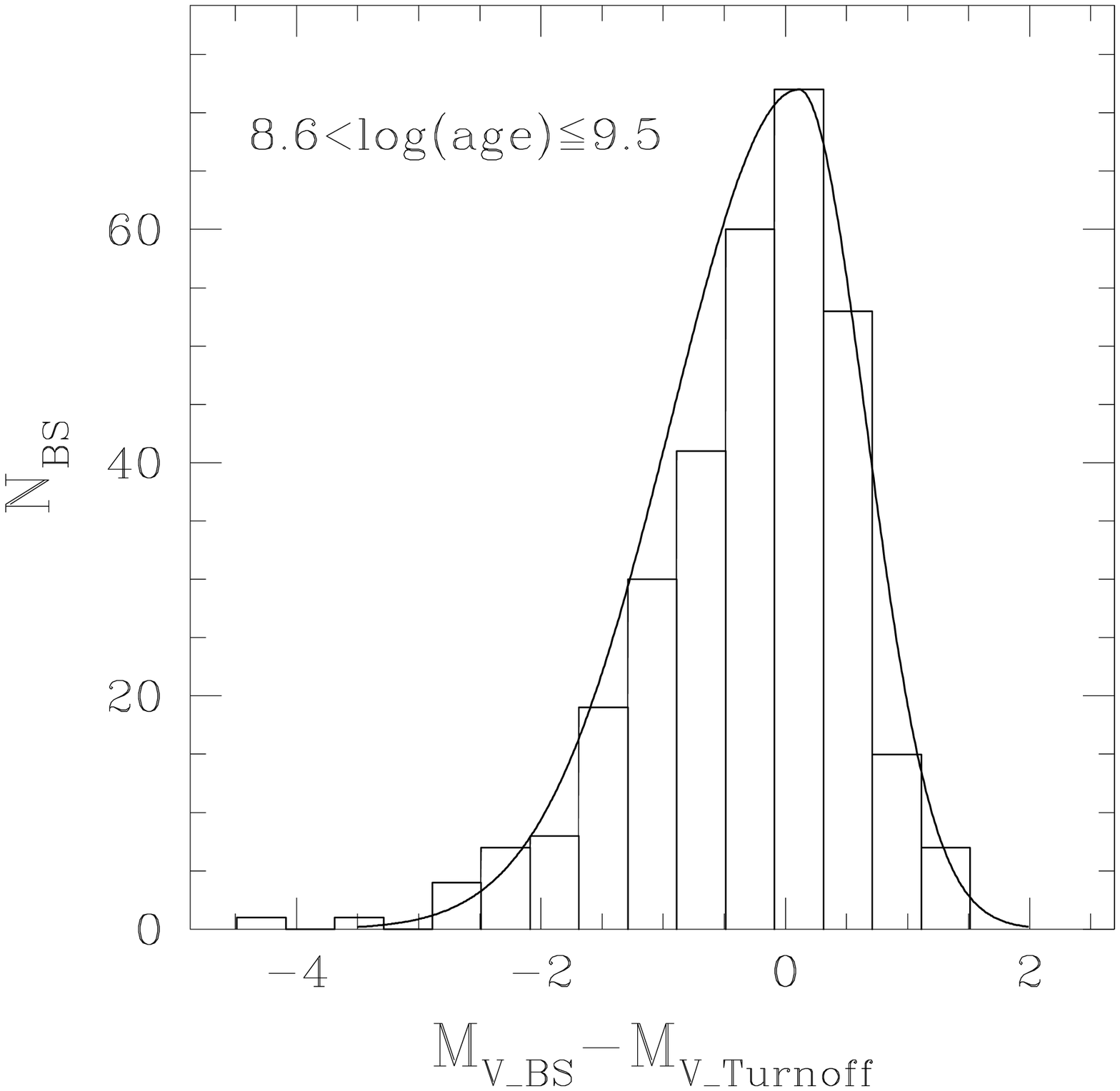}\includegraphics[width=4.3cm]{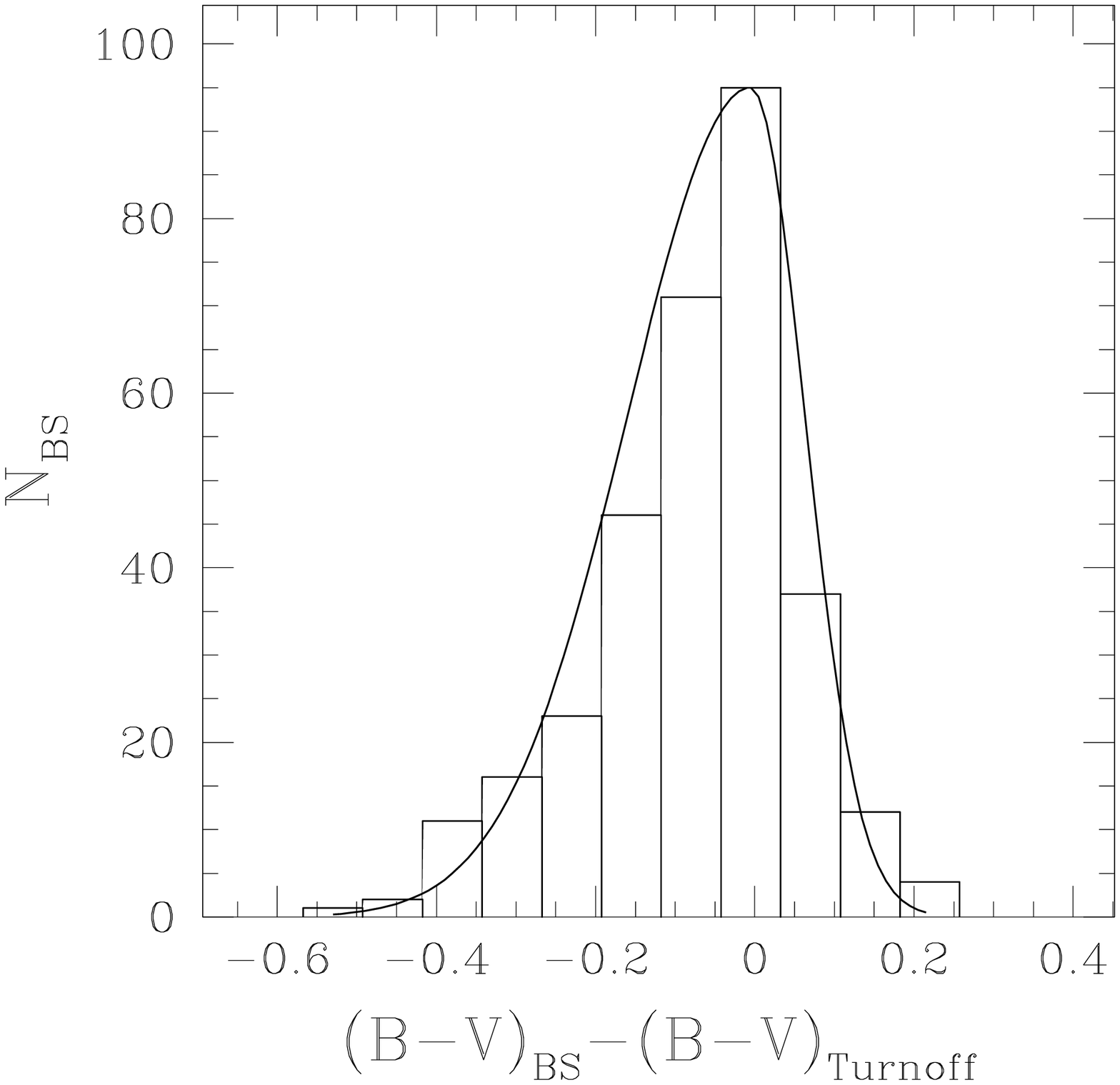}\\
\includegraphics[width=4.3cm]{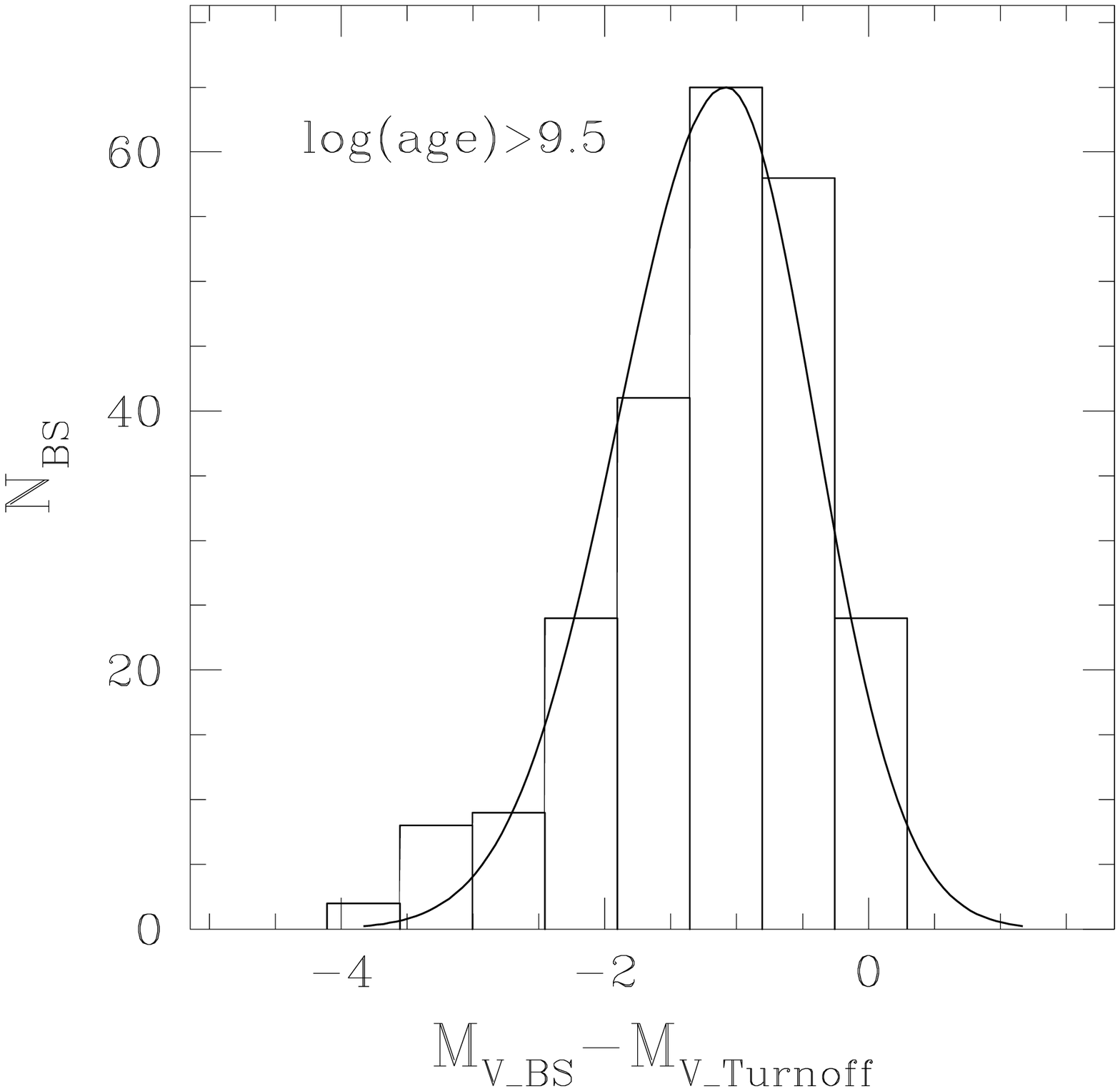}\includegraphics[width=4.3cm]{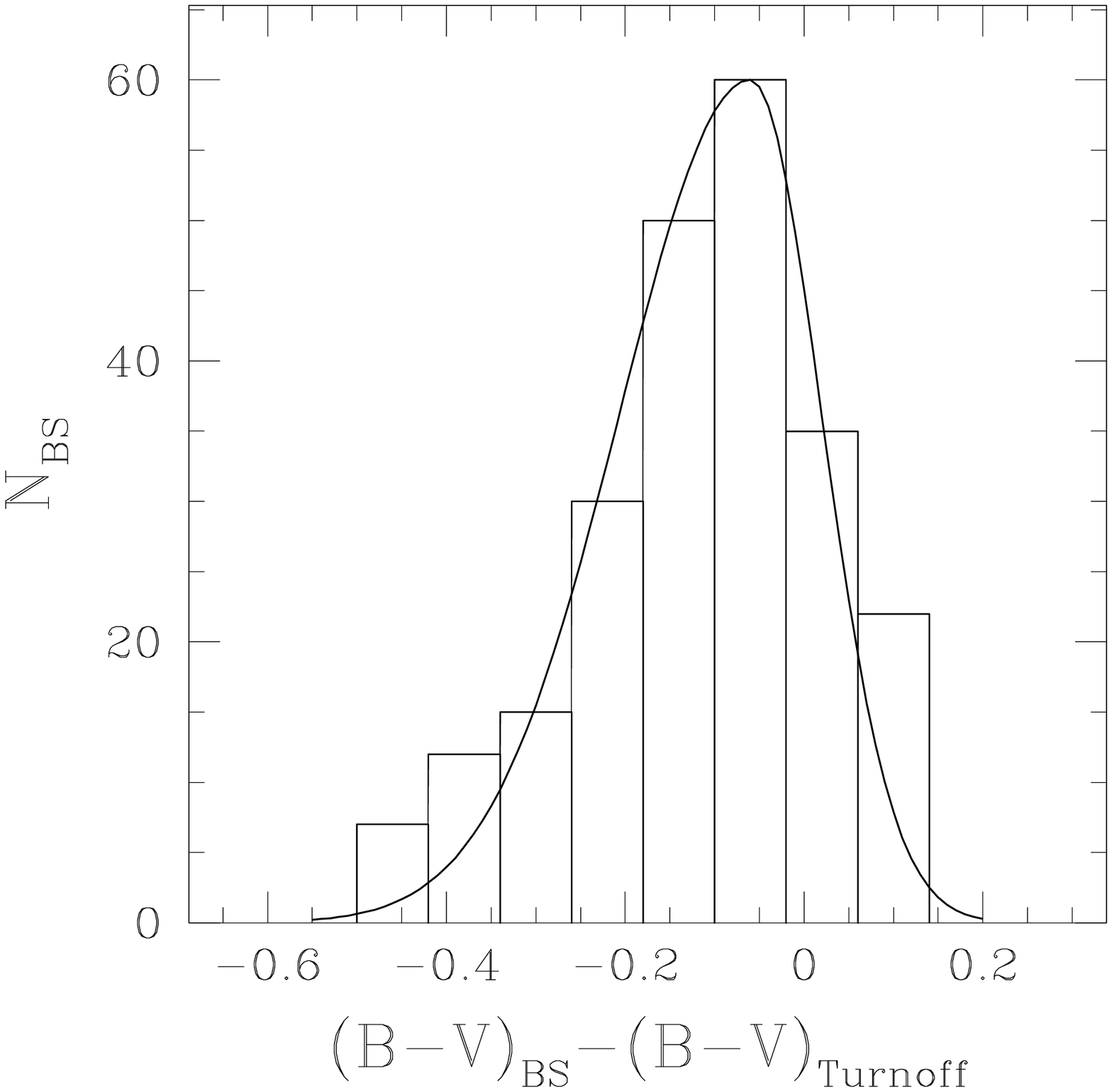}\\
\caption{BS distribution functions in our OC CMDs in three different
  age bins as a function of $M_V$ (left panels) and the $(B-V)$ (right
  panels). In each panel, a Gaussian profile is used to describe
  either side of the peak separately.}
\label{fig5}
\end{figure}

\begin{figure}
\includegraphics[width=4.3cm]{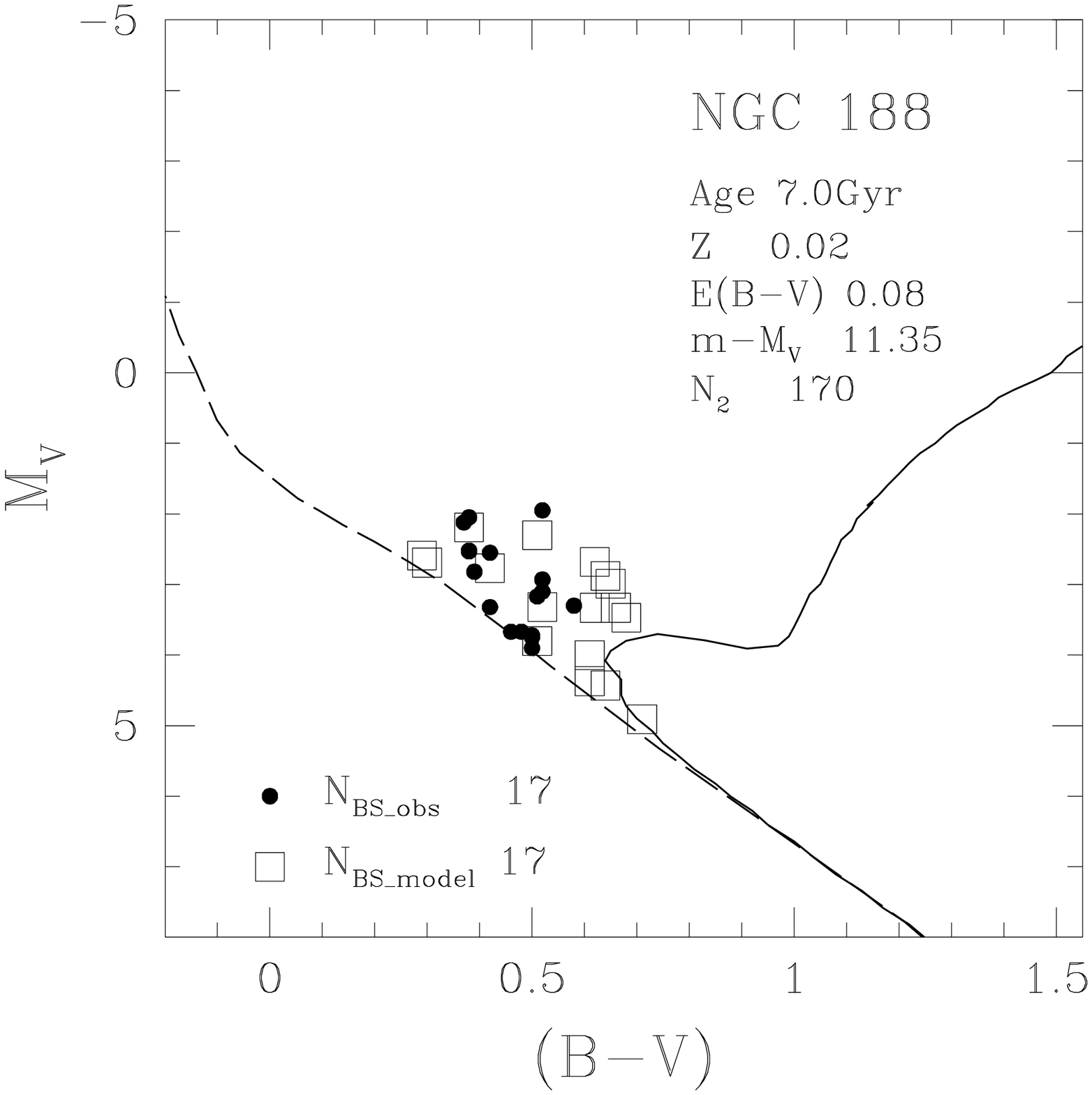}\includegraphics[width=4.3cm]{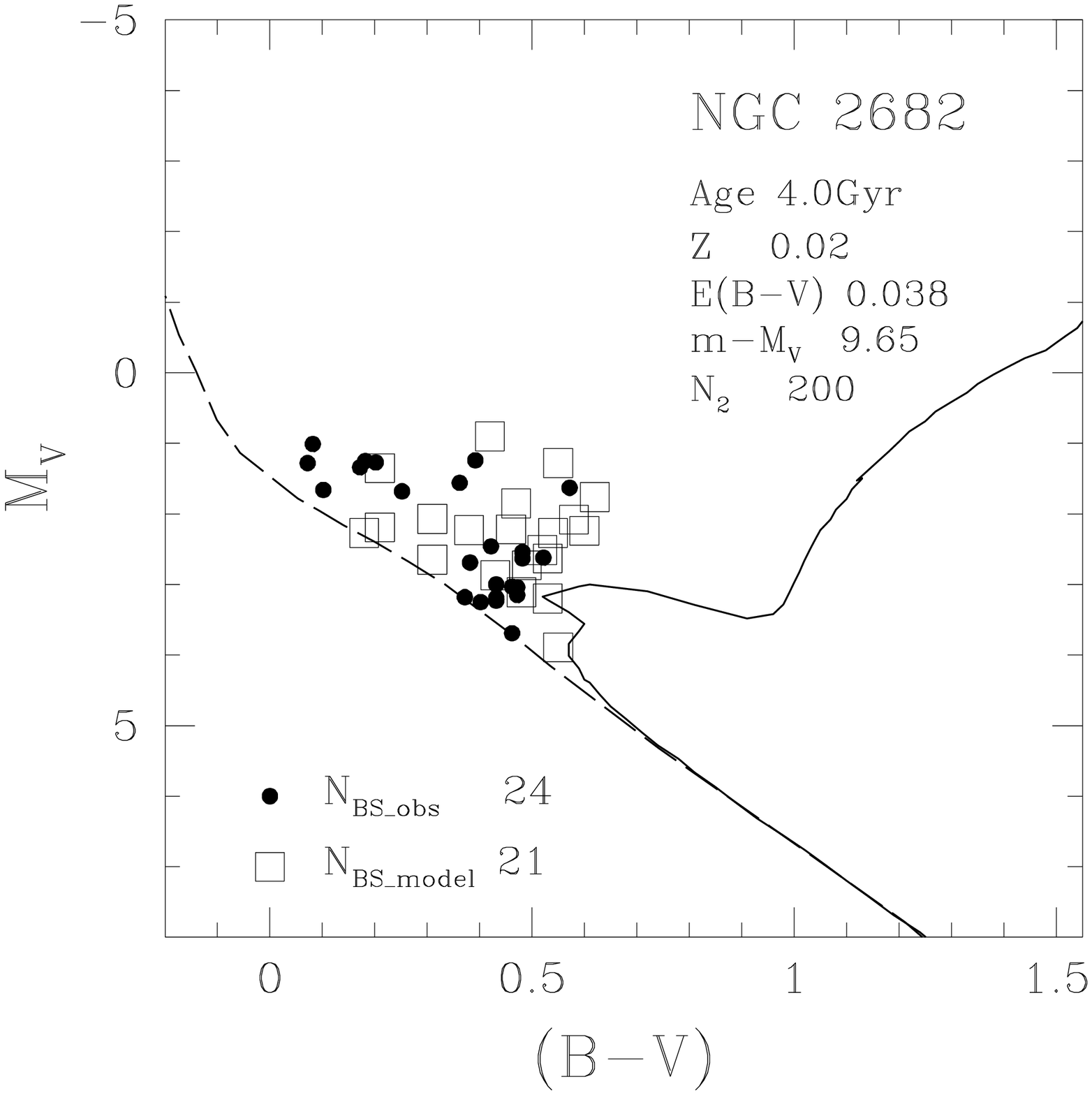}\\
\includegraphics[width=4.3cm]{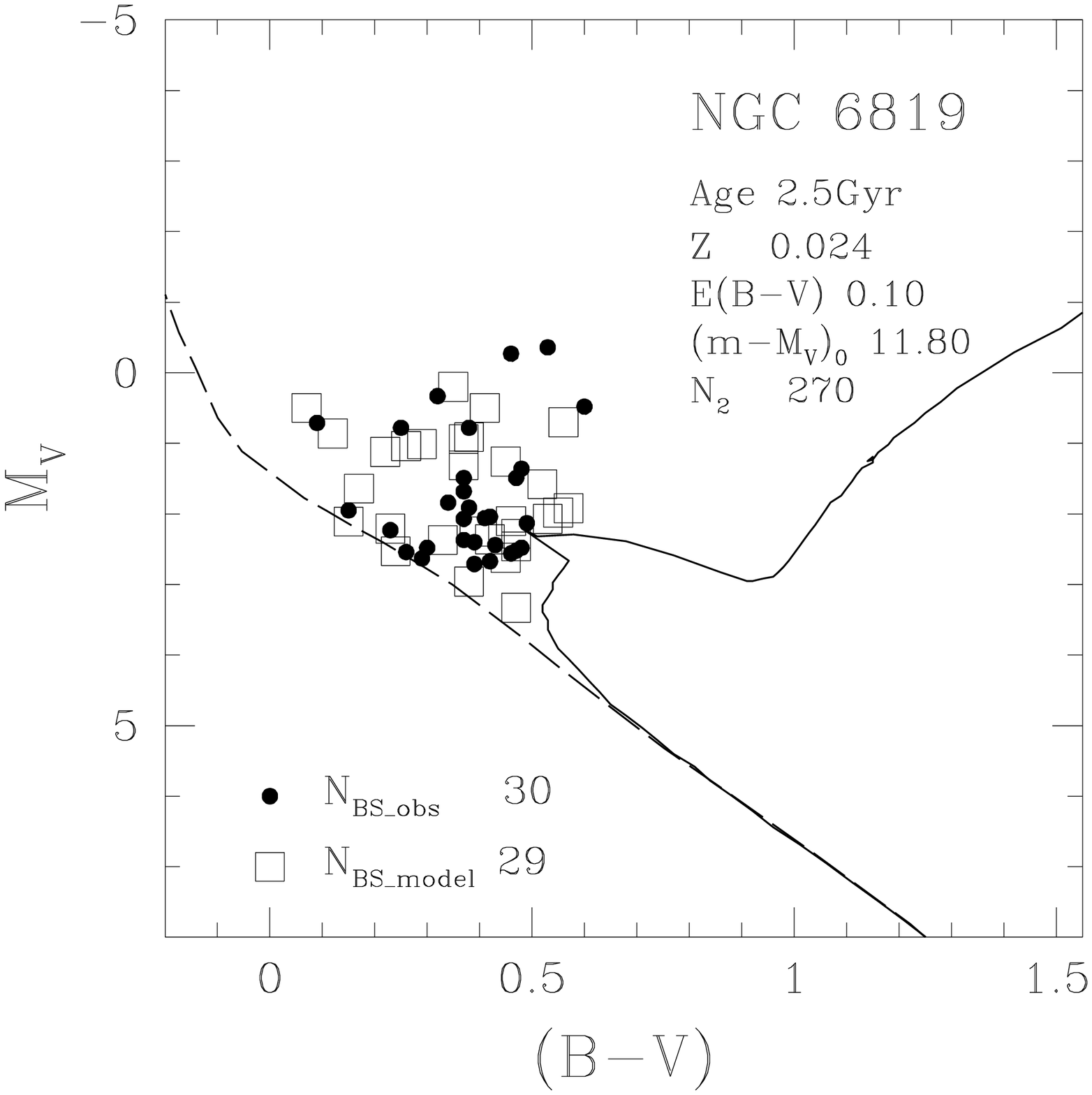}\includegraphics[width=4.3cm]{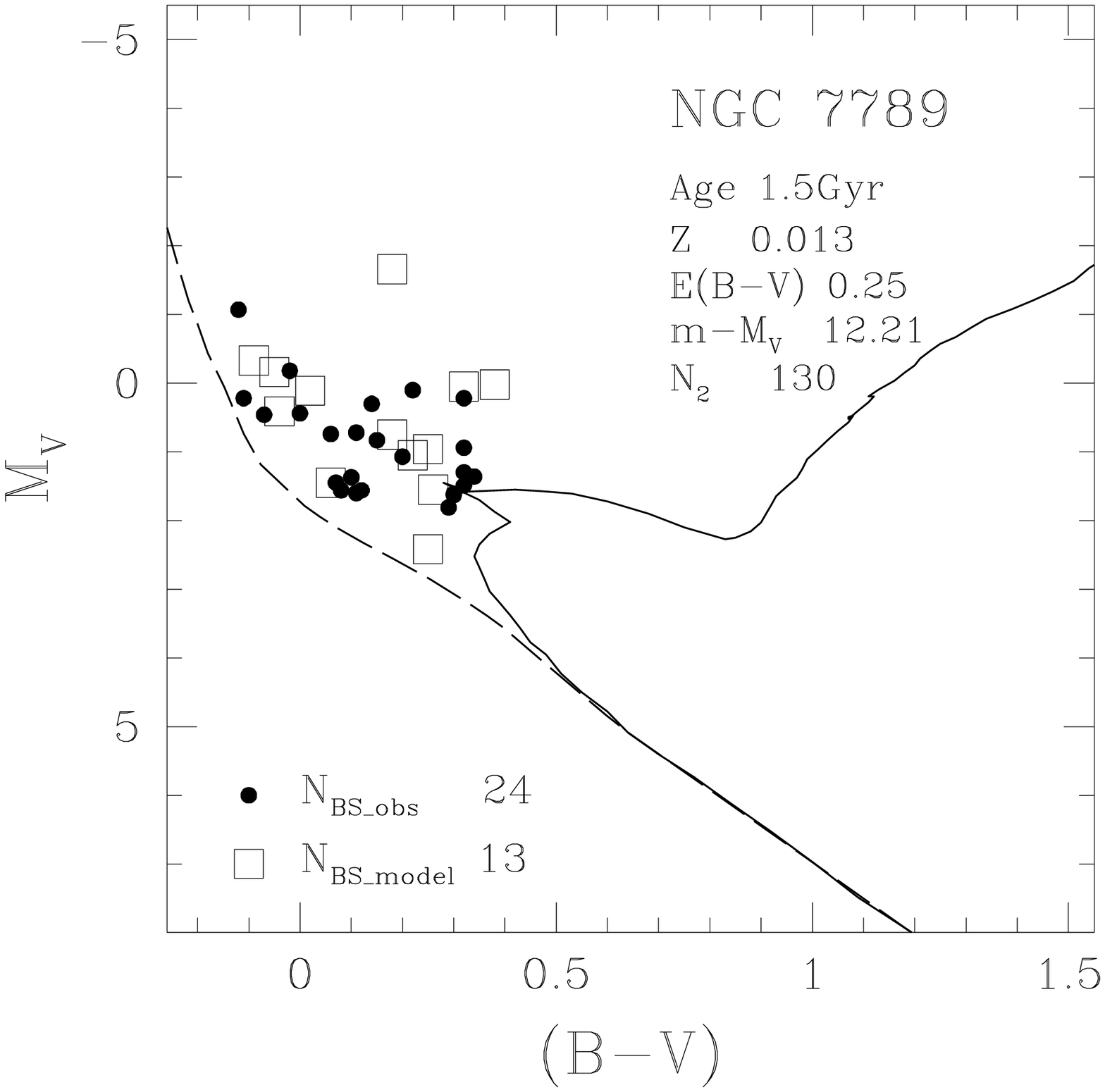}\\
\caption{Comparison between the observed and modelled BS populations
  in the CMDs of four representative Galactic OCs. In each panel, the
  fundamental cluster parameters are given in the top right-hand
  corner, the solid curve is the isochrone representing the cluster's
  age and metallicity, the dashed line is the zero-age MS, the solid
  circles are the observed BSs from AL95, and the open squares are the
  model BSs based on Monte Carlo simulations. $N_{\rm BS\_model}$ is
  calculated using the observed value of $N_2$ in Eq.~(\ref{eq1}).}
\label{fig6}
\end{figure}

The loci of BSs in cluster CMDs are fixed by their formation and
evolutionary processes (Ferraro et al. 2009), which could still be
stochastic owing to the varying physical conditions among star
clusters. Thus, it is impossible to construct the specific
distribution function of the BS population for each individual SSP. A
feasible approach to generate a BS population in the CMD is to work
out a uniform BS distribution function.

The dotted lines in Fig.~\ref{fig4} are isochrones with ages between
0.1 and 20 Gyr, truncated at the bottom of the RGB phase. The two
dashed lines, i.e., the zero-age MS (ZAMS) and the boundary between
the MS and post-MS phases, highlight the entire MS stage for all
isochrones. The distance between the two dashed lines becomes narrower
in the colour range in the CMD for older -- log(age/yr)$\ge$9.5 -- and
younger -- log(age/yr)$\le$8.6 -- SSPs, which means that a uniform BS
distribution function will certainly push a BS population towards red
colours for SSPs with a MS turnoff close to the ZAMS. To avoid
uncertainties associated with the distance between the MS turnoff
points and the ZAMS, we consider BS distribution functions in three
different age bins, i.e., 8.0$\le$log(age/yr)$\le$8.6,
8.6$<$log(age/yr)$\le$9.5, and log(age/yr)$>$9.5. The solid lines in
the figure show the age-bin selection.

Since the bin selection is done empirically, we tested that a small
change in the age-bin selection, i.e., $\Delta \log({\rm
  age/yr})=0.1$, cannot effectively modify the BS distribution in the
CMDs and, consequently, affect our model results. Specifically, we
construct the BS distribution functions with $\log({\rm age/yr})=8.6$
included in the older age bin and calculate the $(U-B)$ and $(B-V)$
colours of the SSP with $Z=0.02$ and $\log({\rm age/yr})=8.6$. The
colour changes resulting from adopting different distribution
functions are $<~0.0001$~mag for both colours. The same result is
found for a test with $\log({\rm age/yr})=9.5$.

We align the MS turnoff points of all OCs in our working sample to
obtain a sufficient number of BSs for good statistics.
Fig.~\ref{fig5} shows the distribution functions of BSs versus the MS
turnoff point in three different age bins as a function of $M_V$ (left
panels) and $(B-V)$ (right panels). For each of the distributions, a
Gaussian profile is adopted to fit either side of the peak separately,
based on which we construct the BS population in the CMD of an SSP
using Monte Carlo simulations in two dimensions ($M_V$ and
$(B-V)$). All Gaussian profiles have the same standard format,
\begin{equation}
f(x)=\exp \left[-\frac{\left(x-\mu\right)^2}{2\sigma^2}\right], \label{eq2}
\end{equation}
where $\mu$ is the peak position and $\sigma$ the standard deviation.
The notation $x=\Delta M_V=M_{V_{\rm BS}}-M_{V_{\rm TO}}$ refers to
the increment in luminosity ($M_V$) of a BS with respect to the SSP's
turnoff luminosity; $x=\Delta (B-V)=(B-V)_{\rm BS}-(B-V)_{\rm TO}$ is
the equivalent increment in colour index $(B-V)$. The specific values
of $\mu$ and $\sigma$ for different $x$ ranges are listed in
Table~\ref{table2}. Here, $f^-(x)$ and $f^+(x)$ refer to the left- and
right-hand sides of the Gaussian profiles, respectively.

Since the distribution functions are constructed empirically from
observational statistics, the values of both $\mu$ and $\sigma$ are
inevitably sensitive to the selection of the bin size. To find
reasonable descriptions for the Gaussian profiles, we start exploring
the statistics of $M_V$ and $(B-V)$ in each age bin with a bin size of
$\Delta M_V=0.20$ and $\Delta (B-V)=0.010$ mag, and we subsequently
increase the bin size in steps of $\Delta M_V=0.05$ and $\Delta
(B-V)=0.005$ mag until the distribution resembles a Gaussian
function. The adopted bin sizes for $\Delta M_V$ and $\Delta (B-V)$
for each age bin are given in column (3) of Table~\ref{table2}. The
$\mu$ and $\sigma$ values listed in the table are calculated based on
the corresponding bin size.

In addition to the distribution function, a further boundary that also
constrains BS positions in the CMD is the ZAMS. Any BSs located
beyond the ZAMS, i.e., with bluer colours than the ZAMS for the same
luminosity, will not be generated by our program. This assumption is
made mainly because we treat BSs as MS stars and describe them using
standard MS models.

\begin{figure*}
\includegraphics[width=5.2cm]{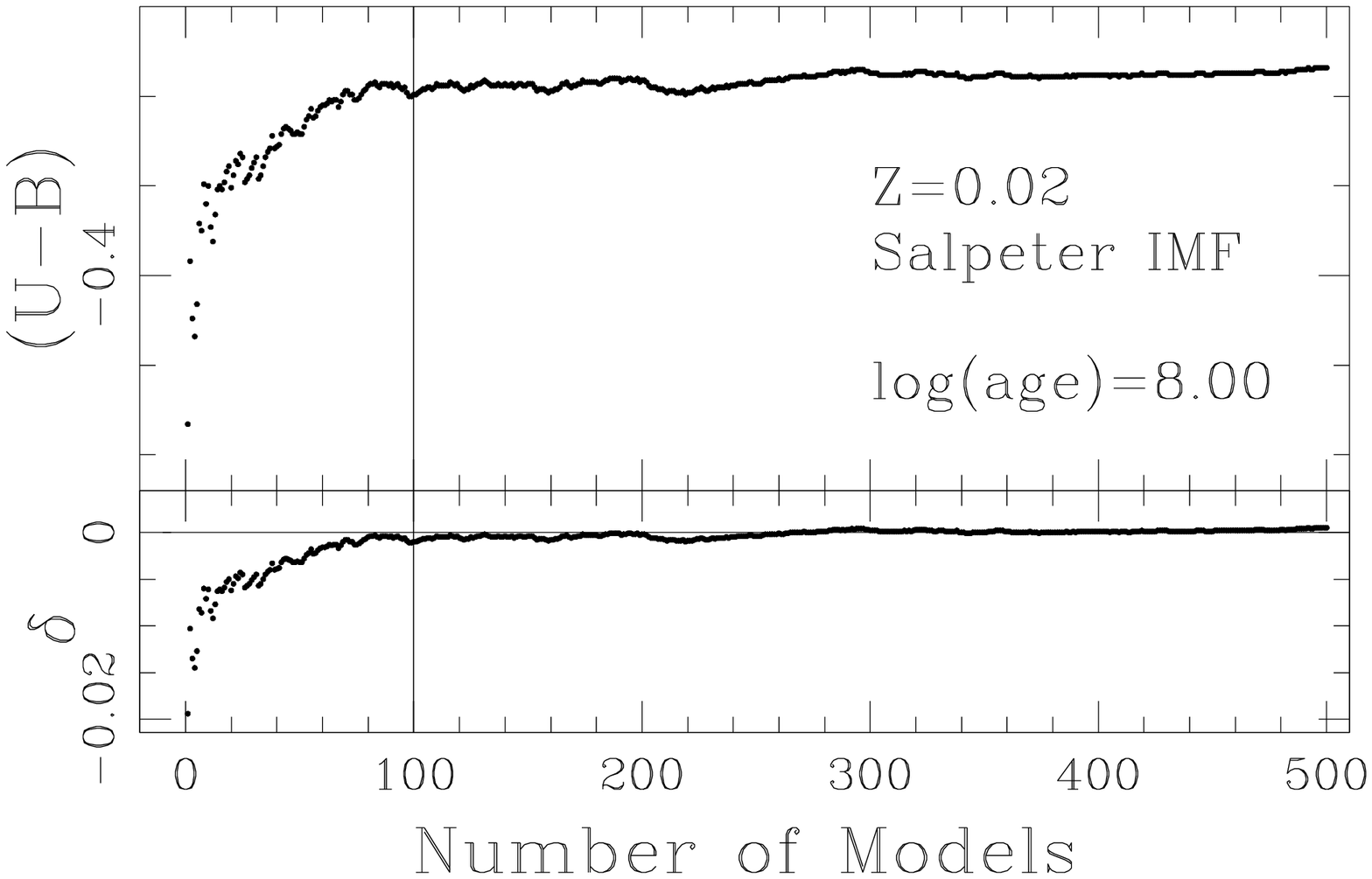}\includegraphics[width=5.2cm]{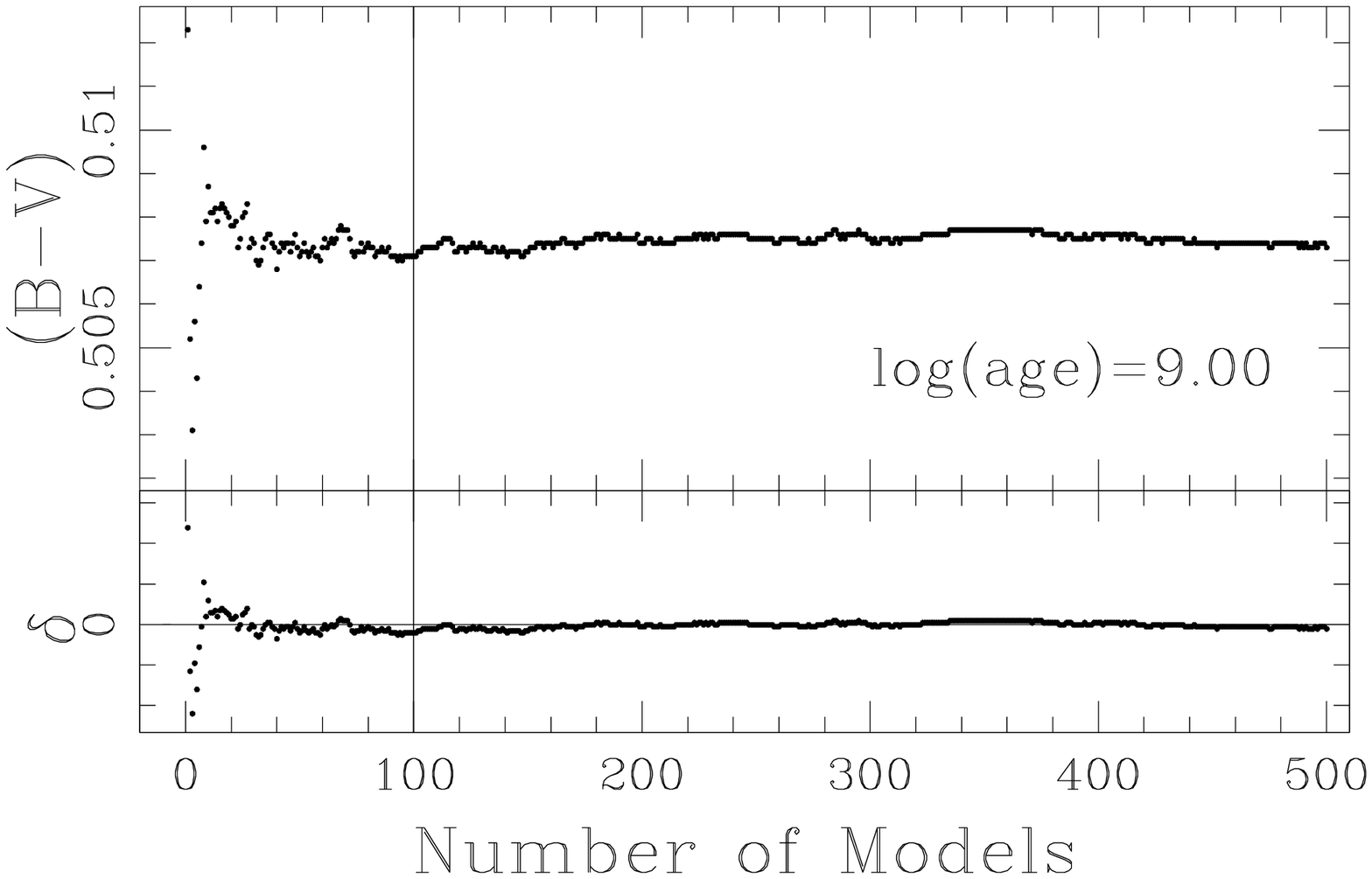}\includegraphics[width=5.2cm]{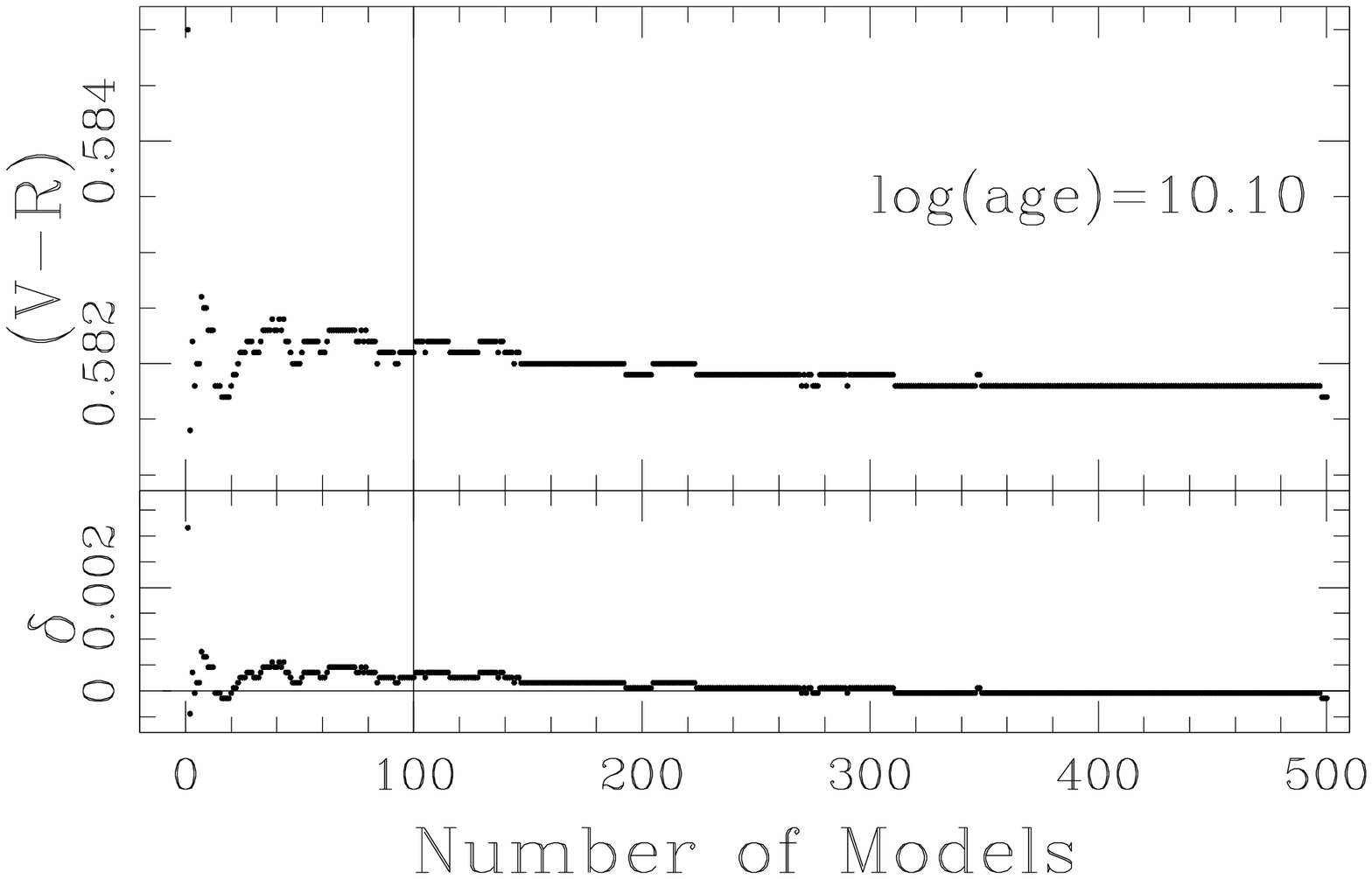}\\
\includegraphics[width=5.2cm]{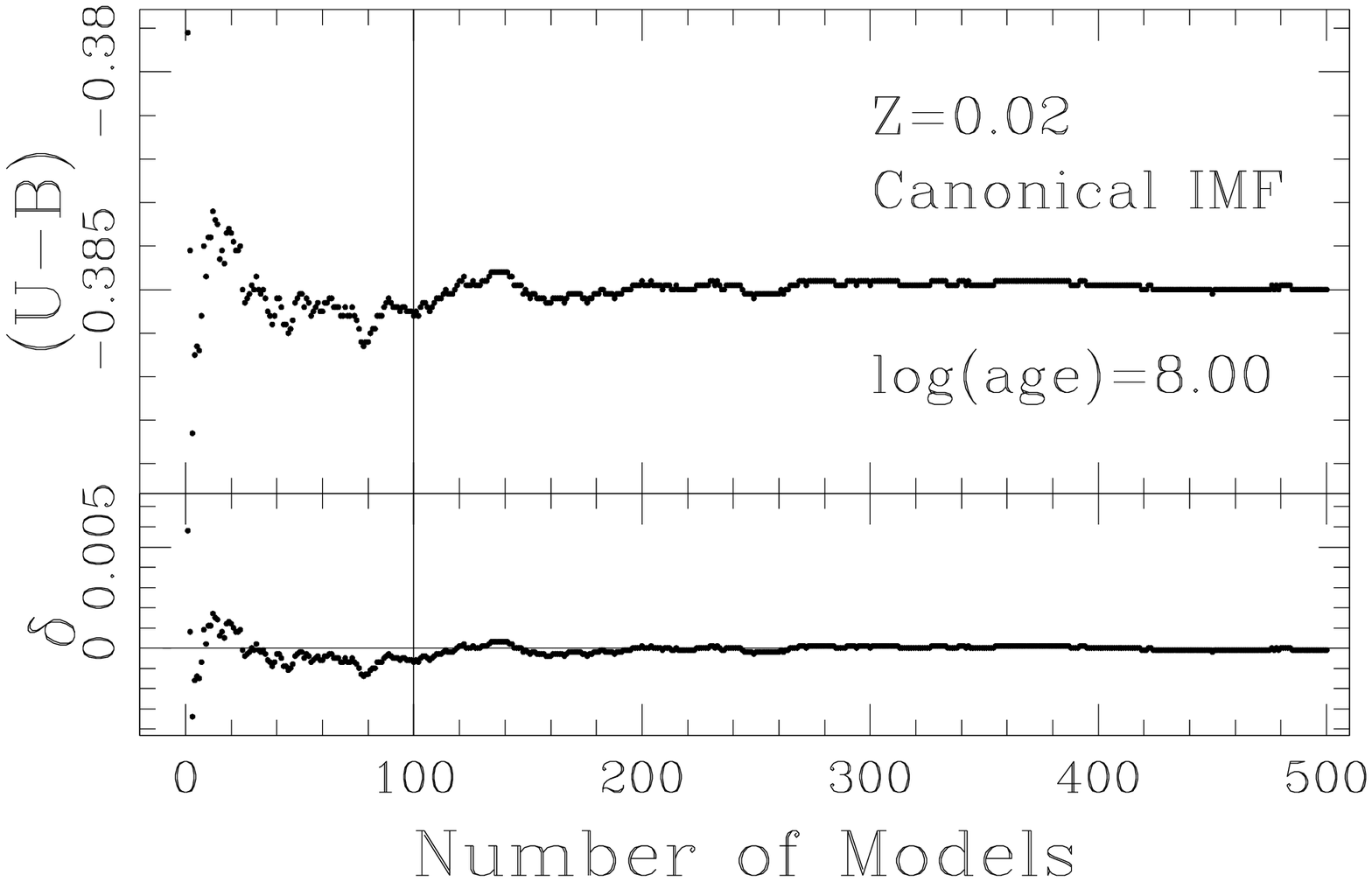}\includegraphics[width=5.2cm]{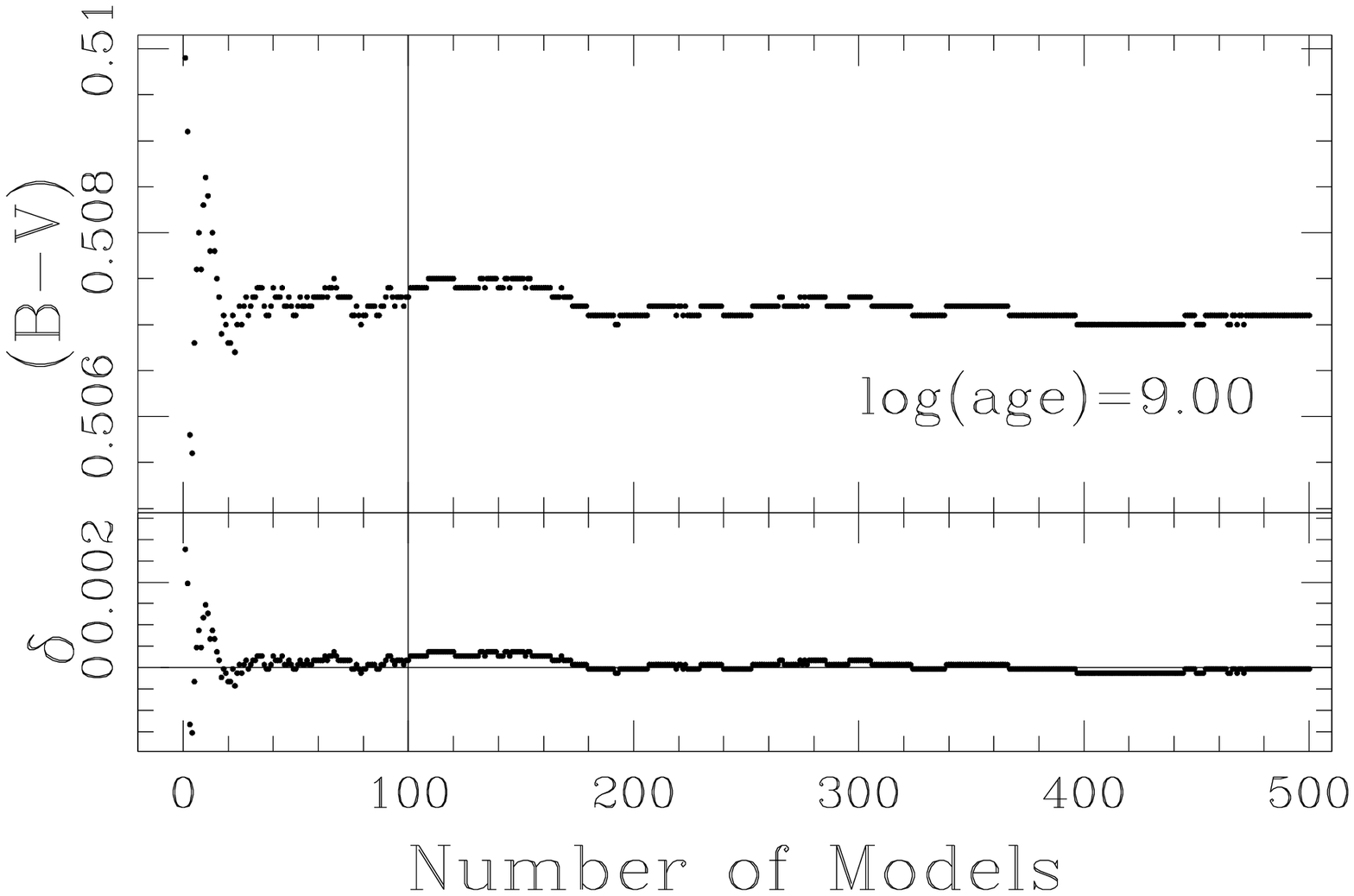}\includegraphics[width=5.2cm]{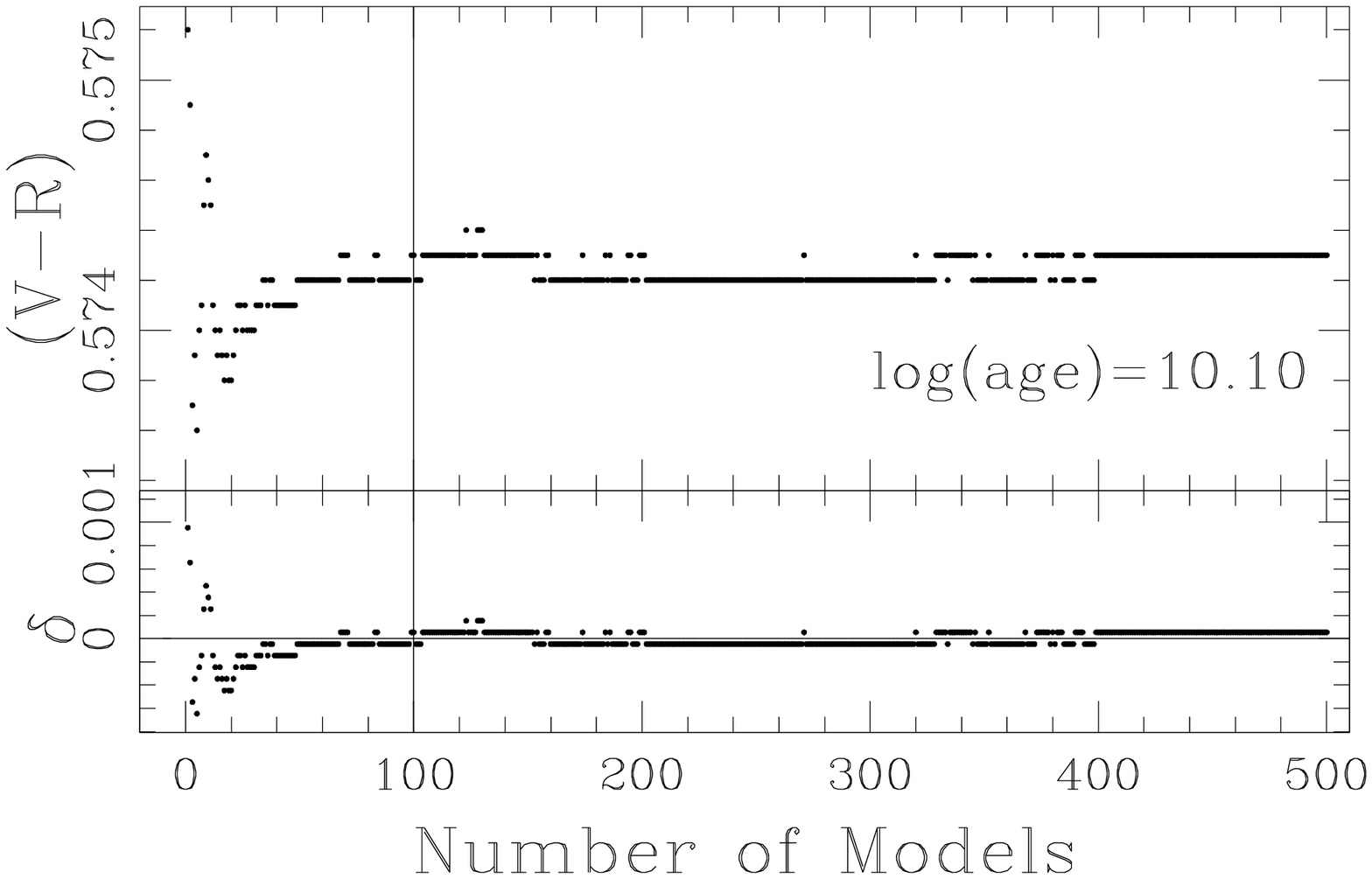}\\
\caption{Choice of the number of models to calculate the spectrum of
  the BS population in our model construction (see Section 2.2). Each
  panel shows a broad-band colour ($y$ axis) versus the number of
  models ($x$ axis) used for calculating the colour. We calculate the
  average of each colour with the colours using between 200 and 500
  models to ensure that the fluctuation is small. On the $y$ axis,
  $\delta$ is the average difference among the colours. Using solar
  metallicity as an example, we search for the number of reliable
  models with a value of $\delta$ $<$ 0.0005~mag for different
  colours, ages and IMFs. Finally, we choose 100 models (marked by the
  vertical solid line in each panel) to construct the spectrum of the
  BS population.}
\label{fig7}
\end{figure*}

\subsection{Model construction}

Our sample of 100 Galactic OCs has limited parameter coverage in age
and metallicity, i.e., it covers ages from 0.1 to 12 Gyr and
metallicities from $Z=0.0048$ to 0.035 (Xin et al. 2007, their table
1). To consider the BS contributions for the full set of SSP models,
some extrapolations of the parameter space have been adopted based on
the results from OCs and some reliable working assumptions: (i) ages
from 0.1 to 20 Gyr, with the lower limit set by the difficulty to
identify BS components in star clusters younger than 100~Myr, while
the BS properties are not expected to change dramatically in very old
stellar populations (see Section 3); and (ii) metallicities $Z=0.0004,
0.004, 0.008, 0.02$ and 0.05. Although the current best
constraint on the age of the Universe is $13.73 \pm 0.12$ Gyr from the
WMAP measurements, we retain an age of 20 Gyr
\footnote{The BC03 SSP models and the stellar evolutionary tracks that
form their basis have been calculated for ages up to 20 Gyr. It is not
straightforward to correct this to a lower age; one would need to
recalculate all stellar evolutionary tracks for all metallicities and,
for instance, update the precise descriptions of opacities,
convection, and evolution of, e.g., horizontal branch (HB) and AGB
stars. This situation is worsened by the fact that young OCs, up to
$\sim5$~Gyr, have been used to properly and robustly constrain stellar
evolution at younger ages, whilst at older ages the observational
diagnostics are less sensitive to changes in age, resulting in
significantly greater uncertainties at those ages. These are therefore
issues that one needs to keep in mind in the context of physical
parameters derived for older GCs using both the BC03 and our models.}
as upper age limit of the models, both to keep the model grids the
same as those of BC03, and to enable the applications of the models to
the studies of stellar populations if a possible correction to the
currently accepted cosmology is required.  We extend the metallicity
to a lower boundary of $Z=0.0004$ and a maximum value of $Z=0.05$
because the statistics of OCs show that BS behaviour is not sensitive
to metallicity. ($Z=0.0001$ is not included in the models because HB
stars, instead of BSs, dominate the energy in the ultraviolet and blue
bands in extremely metal-poor populations.)

Our construction procedure for SSP models including BS contributions
is summarised as follows. (i) We use the standard models (we use BC03
in this paper, but we can in principle use any other SSP flavour as
well) to represent the integrated spectrum of the `normal' SSP member
stars. We subsequently use the statistical properties of BSs from
Galactic OCs to generate the BS population for the appropriate
SSP. (ii) We calculate the spectrum of the BS population and combine
it with the spectrum of the normal member stars after the appropriate
flux calibration. The composite spectrum is the integrated spectrum of
the BS-corrected SSP.

In detail, the model construction includes the following steps.

(1) We assume that any given model SSP contains $10^5$ original member
stars. The corresponding normalisation constant, `$A$', for a given
IMF is calculated as
\begin{equation}
10^5 = A \times \int_{m_{\rm l}}^{m_{\rm u}} \phi(m)~{\rm
  d}m, \label{eq3}
\end{equation}
where $\phi(m)$ is the IMF, $m_{\rm l}=0.1 {\rm M}_{\odot}$ and
$m_{\rm u}=100 {\rm M}_{\odot}$.

(2) The SSP's $N_2$ number is calculated using
\begin{equation}
N_2 = A \times \int_{m_1}^{m_2} \phi(m)~{\rm d}m,  \label{eq4}
\end{equation}
where $m_2$ is the mass of the SSP's MS turnoff point and $m_1$ is the
mass on the MS 2 mag below the turnoff point.

(3) The SSP's $N_{\rm BS}$ then follows from Eq.~(\ref{eq1}). 

The BS catalogue of AL95 and AL07 may suffer from field-star
contamination, as acknowledged by these authors. Without accurate
measurements of the stellar membership probabilities of all OCs, it is
impossible to estimate the field-star influence on our statistics.
However, the way in which we calculate the BS-population energy in
SSPs is designed to reduce these effects. We use the $N_{\rm BS}/N_2$
ratio, instead of just $N_{\rm BS}$, to obtain the `relative'
BS-population energy with respect to the energy of a classical SSP,
which is controlled by $N_2$. Given the general make-up of galactic
fields, there are significantly fewer luminous stars compared to the
number of MS stars, so that our adoption of $N_{\rm BS}/N_2$
facilitates maintenance of the BS contributions in SSPs close to the
real situation.

(4) The distribution of the BS population in the CMD is generated
using Monte Carlo simulations in two dimensions in the CMD, i.e.,
$M_V$ and $(B-V)$, based on the Gaussian profiles described by
Eq.~(\ref{eq2}) and Table~\ref{table2}.

Fig.~\ref{fig6} shows the comparison between the observed and modelled
BS populations in the CMDs of four representative Galactic OCs. In
each panel, the fundamental cluster parameters are listed in the top
right-hand corner, the solid curve is the Padova1994 isochrone for the
cluster's age and metallicity, the dashed line is the ZAMS, the solid
circles are the observed BSs from AL95 and the open squares are the
BSs generated by the Monte Carlo simulation. We calculate $N_{\rm
  BS\_model}$ from the observed $N_2$ and Eq.~(\ref{eq1}). Apparently,
the BS distribution fluctuates significantly for different simulations
because of the small $N_{\rm BS\_model}$ numbers. What we intend to
show with this figure is that the modelled BS population is quite
reasonable and comparable with that observed.

(5) Calculate the spectrum of the SSP's BS population.

The distribution of BS loci in the CMDs can influence the
stability of the spectrum of the BS population. One of the best ways
to reduce this stochastic effect is by using the average of a large
number of models. To find the optimum number of realisations, (i) we
repeat the generation of BS populations in the CMD for a given SSP 500
times (500 models) and we subsequently calculate the corresponding
composite spectrum of the SSP. This implies that the 500 models yield
500 different spectra for the same SSP. (ii) We calculate the
broad-band colours $(U-B)$, $(B-V)$ and $(V-R)$ for the SSP based on
the average of the spectra of a successively increasing number till
500. This calculation results in 500 values for each colour. (iii) We
calculate the average of each colour using the colour values for
between 200 and 500 models, thus quantifying the differences
($\delta$) of the 500 colour values and the average colour. Finally,
(iv) we identify the number of models adopted to repeat generating the
BS population in the CMD as the number for which $\delta <
0.0005$~mag. Using the solar-metallicity SSPs as templates, we
conclude that combining 100 models is a safe choice for SSPs
characterised by different ages and IMFs. Direct impressions of the
changes in colours for different numbers of models and positions for
100 models are presented in Fig.~\ref{fig7}.

\begin{figure}
\begin{center}
\includegraphics[width=8cm]{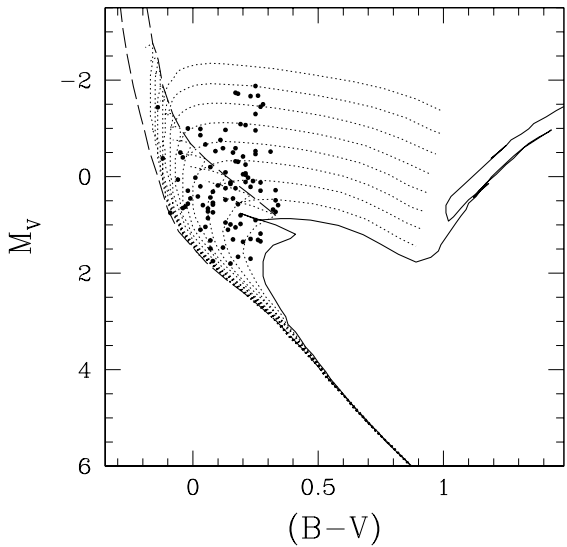}
\caption{Derivation of the basic parameters of BSs in the CMD. The
  solid dots are the model BSs, the solid curve is the isochrone for
  the SSP's age and metallicity, and the dotted lines are isochrones
  representing ages younger than the SSP and truncated at the bottom
  of the RGB phase. The dashed lines are the ZAMS and boundary between
  the MS and post-MS stages, respectively.}
\label{fig8}
\end{center}
\end{figure}

To construct the spectrum of the model BS population, we use
Padova1994 isochrones of the same metallicity but younger ages than
the SSP to fit the position of each BS in the CMD. We then derive the
effective temperature ($T_{\rm eff}$) and surface gravity (log~$g$) by
interpolation between two isochrones straddling the BS. A
demonstration of this procedure is included in Fig.~\ref{fig8}. The
solid dots are the model BSs, the solid line is the isochrone for the
SSP's age and metallicity and the dotted lines are isochrones with
ages younger than the SSP and truncated at the bottom of the RGB
phase. The dashed lines are the ZAMS and the boundary of the MS and
post-MS stages, respectively. In our model construction, BSs located
between the dashed lines are modelled strictly assuming that they can
be represented by the MS phases of the isochrones and the remainder of
the BSs located outside the boundary are fitted with post-MS (and
pre-RGB) phases.

Depending on the values of $T_{\rm eff}$ and log~$g$, a spectrum is
extracted from the Lejeune et al. (1997) spectral library and assigned
to the BS. The flux of the BS spectrum is then calibrated using the
BS's absolute magnitude and, finally, the spectrum of the BS
population is obtained by adding up all flux-calibrated BS spectra,
i.e., $F_{\rm BS} = \sum_{i=1}^{N_{\rm BS}}f_{\rm BS}^i$.

The approximation to use the spectra of single MS stars to represent
the spectra of BSs is made based on both theoretical (e.g., Benz \&
Hills 1987, 1992) and observational (e.g., Shetrone \& Sandquist 2000;
Liu et al. 2008) considerations. The major formation scenarios of BSs,
such as mergers of primordial binaries and dynamical encounters
between stars, can replenish fresh hydrogen fuel in the core and
rejuvenate BSs to the MS stage. Liu et al. (2008) studied the spectral
properties of a complete sample of 24 BSs in the Galactic OC NGC~2682
(M67) based on spectroscopic observations with a resolution of
3.2{\AA}~pixel$^{-1}$ and covering wavelengths of 3600--6900
{\AA}. They concluded that BS spectra can be well represented by the
theoretical spectra of single stars, at least at medium resolution.

(6) We use BC03 models to represent the spectrum of the population of
normal member stars in an SSP (i.e., all member stars except the BSs).

For a conventional SSP of age $t$ and metallicity $Z$, the integrated
spectrum is given by
\begin{equation}
 F_{\rm SSP} (\lambda, t, Z) = B \times \int_{m_{\rm l}}^{m_{\rm u}}
 \phi(m) f(\lambda, m, t, Z)~{\rm d}m, \label{eq5}
\end{equation}
where $\phi(m)$ is the IMF, $f(\lambda, m, t, Z)$ is the spectrum of a
single star of mass $m$, age $t$ and metallicity $Z$, $m_{\rm u}$ and
$m_{\rm l}$ are the upper and lower integration limits in mass,
respectively, and `$B$' is the normalisation constant required to
restore the real intensity of the flux of the conventional SSP.

Since BC03 normalised the total mass of their model SSPs to
1~M$_{\odot}$ at $t=0$, `$B$' is the total mass of the model SSP
containing $10^5$ stars at $t=0$:
\begin{equation}
 M_{\rm tot} = B = A \times \int_{m_1}^{m_2} \phi(m) m~{\rm d}m, \label{eq6}
\end{equation}
where `$A$' is the normalisation constant from Eq.~(\ref{eq3}), $m_1
=0.1$ and $m_2=100 {\rm M}_{\odot}$.

Similarly as for the calibration of the BS spectra, the flux of the
conventional SSP is also calibrated based on its absolute
magnitude. $M_{\rm tot}$ is used to calculate $M_V$ of the
conventional SSP ($M_{V\_{\rm SSP}}$) based on Eq.~(\ref{eq7}), in
which $M_{V\_ {\rm SSP(BC03)}}$ is the absolute magnitude of the
corresponding BC03 SSP. ($M_{\rm tot}$ actually quantifies the
relative increase in $M_V$ with respect to the $M_V$ of an initially $1M_{\odot}$ SSP.)

\begin{equation}
M_{V\_{\rm SSP}} = M_{V\_{\rm SSP(BC03)}} + 2.5 \times \log_{10}
(1/M_{\rm tot}) \label{eq7}
\end{equation}

(7) After flux calibration, direct combination of the spectra of the
BS population and the conventional SSP yields the spectrum of the
BS-corrected SSP. As our final step, the composite spectra are
normalised to the BC03 models by adopting the flux-calibration
constant derived from Eq.~(\ref{eq7}) for each SSP.
\begin{figure*}
\begin{center}
\includegraphics[width=12cm]{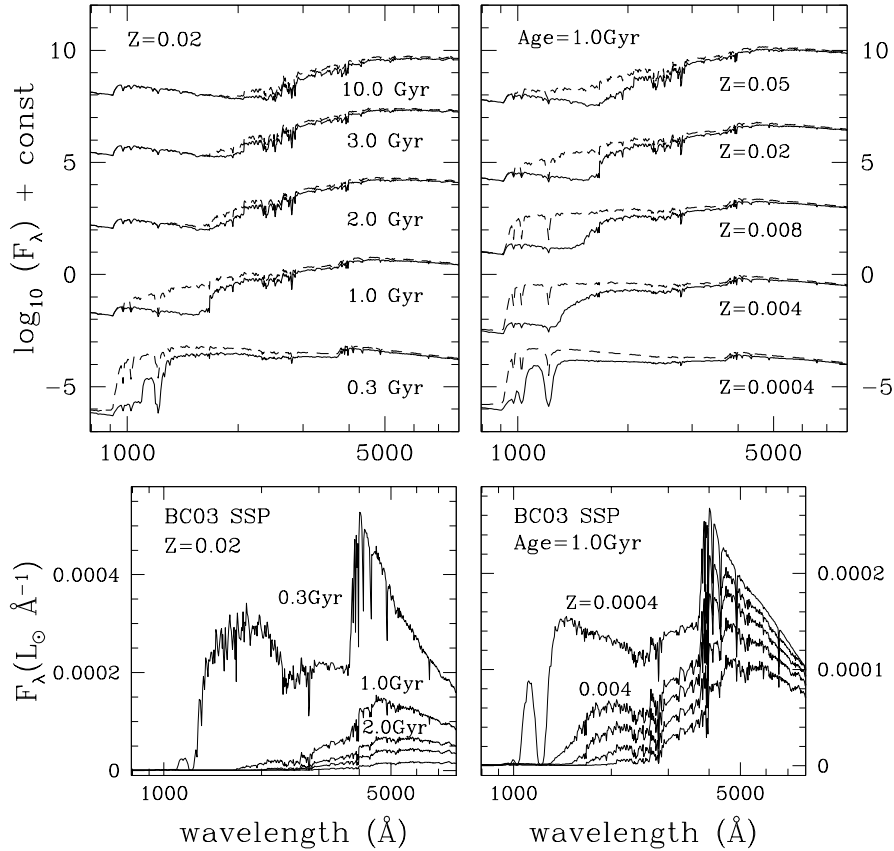}
\caption{Differences of the integrated spectral energy distributions
  (ISEDs) between BC03 and our models for different ages (top
  left-hand panel) and metallicities (top right-hand panel). The solid
  lines in the top panels are the BC03 ISEDs, while the dashed lines
  are the ISEDs resulting from our models. Because the BS contribution
  is calculated as an increment to the BC03 models, the corresponding
  BC03 ISEDs are shown in the bottom panels for reference.}
\label{fig9}
\end{center}
\end{figure*}

\section{Model results}

As an SSP ages, the SSP's BS population evolves to redder colours
in the CMD following the movement of the SSP's MS turnoff, which
implies that the `blue' in `blue stragglers' only means `bluer than
the MS turnoff', not necessarily blue in colour. Since young SSPs
(age$<$0.1 Gyr) are not included in our models, and because of the
presence of `yellow stragglers' in the AL95 catalogue, we decided to
explore the BS contributions to the integrated-light properties
involving $U$-, $B$- and $V$-band energies, because they may all be
significant. Specifically, we present and discuss in detail the
differences in the integrated spectral energy distributions (ISEDs),
broad-band colours and mass-to-light ratios ($M/L_V$) between our
models and those published by BC03 in this section.

Using the $Z=0.02$ models with a Salpeter IMF as example, the
differences in the ISEDs between two models of different age (top
left-hand panel) and metallicity (top right-hand panel) are given in
Fig.~\ref{fig9}. The solid lines in the top panels represent the BC03
ISEDs, while the dashed lines represent the ISEDs from our
models. Since the BS contribution is calculated as an increment to the
BC03 models, the intensities of the corresponding BC03 ISEDs of
different age (bottom left-hand panel) and metallicity (bottom
right-hand panel) are also given in the figure.

The differences presented in the top left-hand panel show a tendency
for a stronger BS contribution for younger SSPs. BSs have a
significant effect on ISEDs for ages between 0.1 and 1.0 Gyr. A sharp
enhancement appears in the UV range at 1.0 Gyr for our models, while
the UV intensity remains very low in the BC03 ISEDs. During SSP
evolution, all massive single stars have left the MS and evolved into
the red supergiant or the red giant phases in conventional SSPs of
0.1--1.0 Gyr, the UV light declines and the near-infrared intensity
increases, and thus BSs are the most luminous and bluest objects in
the populations. For SSPs older than 2.0 Gyr, the intensity of the BS
contribution decreases smoothly and slowly as the population ages.
The bottom left-hand panel presents the BC03 ISEDs for different ages,
which show that ISEDs are stronger (brighter) for younger SSPs. This
tendency is consistent with that found for the BS contribution to
ISEDs of different ages. 

The BS contributions to the ISEDs of SSPs resulting from our
models are consistent, in terms of time scales, with the results of
Chen \& Han (2009).  These authors studied the generation of BSs
produced from primordial binaries and concluded that such BSs, without
angular-momentum loss, can have a strong effect on the resulting ISEDs
in UV and blue bands between 0.3 and 2.0 Gyr (their fig. 4).  Since
their calculations of BS evolution stop before the `blue stragglers'
evolve into the `yellow stragglers' (their fig. 1), we cannot trace
any significant BS contributions to the optical spectral range.  Given
their results, a time scale of 0.1 Gyr is not long enough for BSs
formed in primordial binaries to evolve to loci brighter than the MS
turnoff. This also supports the choice of the lower age limit in our
models.

We use the models for 1.0 Gyr with a Salpeter IMF as example to
highlight the differences between ISEDs of different metallicities
(top right-hand panel in Fig.~\ref{fig9}). The results show that the
BS contribution decreases for more metal-rich SSPs. This tendency is
also consistent with the changes in the BC03 ISEDs of different
metallicities (bottom right-hand panel).

\begin{figure}
\includegraphics[width=8.5cm]{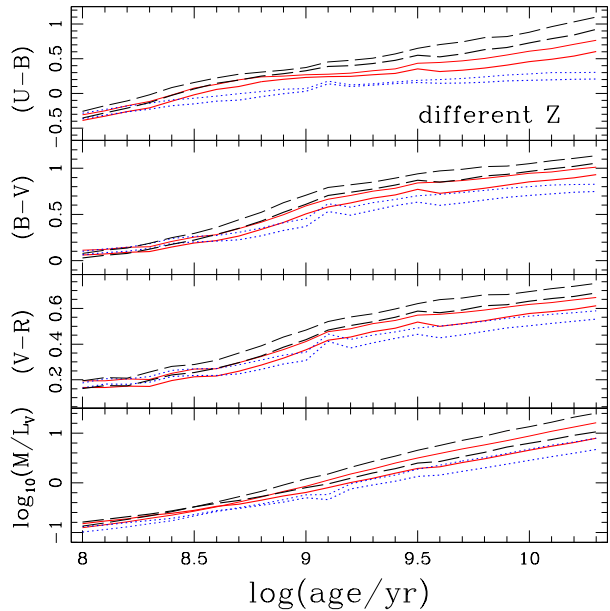}
\caption{Comparison of BC03 and our models for the $(U-B)$, $(B-V)$
  and $(V-R)$ colours and the mass-to-light ratio ($M/L_V$) for three
  different metallicities, $Z=0.004$ (dotted lines), 0.02 (solid
  lines) and 0.05 (dashed lines). The heavy lines represent our
  models. Values shown in the figure are from models with a Salpeter
  IMF.}
\label{fig10}
\end{figure}

Fig.~\ref{fig10} shows the differences in $(U-B)$, $(B-V)$ and $(V-R)$
colours and in the mass-to-light ratio ($\log_{10}(M/L_V)$) between
two models constructed using a Salpeter IMF. A comparison is given for
three different metallicities, $Z=0.004$ (dotted lines), 0.02 (solid
lines) and 0.05 (dashed lines). The heavy lines represent the values
from our models, while the regular lines show the BC03
results. Table~\ref{table3} includes some examples of broad-band
colours and $\log_{10}(M/L_V)$ of BC03 and our models for different
ages, metallicities and IMFs.

To roughly estimate the BS-induced colour modification, we calculate
the average of the colour differences between two models of $Z=0.02$
in the age range from 0.1 to 20 Gyr. These differences are
0.10$\pm$0.05 mag for $(U-B)$, 0.08$\pm$0.02 mag for $(B-V)$ and
0.05$\pm$0.01 mag for $(V-R)$. Our models are characterised by bluer
colours. Such colour modifications are far beyond (or at least
comparable) to the uncertainties inherent to most of the current
generation of advanced observational instrumentation. This means that
our models will be able to contribute a systematic and significant
shift in colour in the direction implying that the old population is
actually older, or more metal-rich, compared to results from
conventional SSP models. This will potentially have important
consequences for the corresponding interpretation regarding galactic
formation and evolution histories.

A comparison of the $M/L_V$ ratio between two models is given by
assuming mass conservation during BS formation, i.e., BS formation
would not change the total mass of a standard SSP but only increase
its total luminosity. The average modification, $\Delta
\log_{10}(M/L_V) = 0.16\pm0.15$ dex, as shown in the lowest panel of
Fig.~\ref{fig10}, is equivalent to $\sim$30\% enhancement in
luminosity ($L_V$) for a given SSP. The enhancement is similar for all
metallicities.

\begin{figure}
\includegraphics[width=8.5cm]{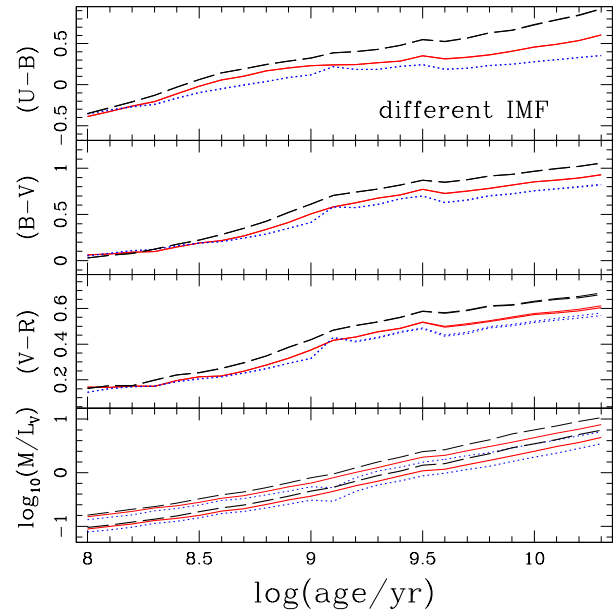}
\caption{Evolution of the $(U-B)$, $(B-V)$ and $(V-R)$ colours and the
  mass-to-light ratio of our models based on different IMFs, i.e., the
  Salpeter and Canonical IMFs (heavy lines). Different line styles
  correspond to different metallicities, i.e., $Z=0.008$ (dotted
  lines), $Z=0.02$ (solid lines) and $Z=0.05$ (dashed lines).}
\label{fig11}
\end{figure}

Fig.~\ref{fig11} shows the evolution of the $(U-B)$, $(B-V)$ and
$(V-R)$ colours and the mass-to-light ratio of our models constructed
with different IMFs. The regular lines represent the models governed
by a Salpeter IMF. The heavy lines are the models constructed with the
Canonical IMF. Different line styles correspond to different
metallicities, i.e., $Z=0.008$ (dotted lines), $Z=0.02$ (solid lines)
and $Z=0.05$ (dashed lines). The evolution of the broad-band colours
does not depend sensitively on the IMF. On the other hand, the
$\log_{10}(M/L_V)$ values are strongly sensitive to the IMF. The
average difference in $\log_{10}(M/L_V)$ between the models of
different IMFs is $\sim0.30\pm0.02$ dex. The Canonical IMF (heavy
lines) causes a stronger effect on $M/L_V$ mainly because it produces
a higher $N_{\rm BS}$ number than the Salpeter IMF for a given SSP.

\begin{table*}
\caption{Examples of broad-band colours and $M/L_V$ for BC03 and our
  models for different ages and metallicities.}
\label{table3}
\begin{tabular}{|cc|ccccc|ccccc|}
\hline
\hline
Age & $Z$& $(U-B)$ & $(B-V)$ & $(V-R)$ & $(V-I)$& $\log_{10}(M/L_V)$ & $(U-B)$ & $(B-V)$ & $(V-R)$ &$(V-I)$ & $\log_{10}(M/L_V)$ \\
(Gyr)& & \multicolumn{5}{c}{BC03 models with the Salpeter IMF} &\multicolumn{5}{c}{Our models with the Salpeter IMF}\\
\hline
      &0.0004& $-$0.1412 & 0.1950 & 0.2527 & 0.4627 & 0.3660 & $-$0.2127 &  0.1343 &  0.2058 &  0.3831 & 0.2997\\
      & 0.004& $-$0.0068 & 0.3072 & 0.2850 & 0.5850 & 0.3750 & $-$0.0980 &  0.2243 &  0.2337 & 0.4910 & 0.3023\\
0.5  &0.008 & 0.0781 & 0.3225 & 0.2875 & 0.5914 & 0.4111 &  $-$0.0088 &  0.2426 &  0.2375 &  0.4989 & 0.3294\\
      & 0.02 & 0.1951 & 0.3463 & 0.2966  & 0.6020 & 0.4695 & 0.1005 &  0.2672 &  0.2478 & 0.5151 & 0.3706 \\
      & 0.05 & 0.2772 & 0.4424 & 0.3504 &0.6870 &  0.5483 & 0.1933 &  0.3503 &  0.2962 & 0.5899 &0.4383 \\
\hline
      &0.0004& $-$0.1370 & 0.3601 & 0.3354 & 0.6179 & 0.5424 & $-$0.1741 &  0.2754 &  0.2829 &  0.5309 &  0.4499\\
      & 0.004 & 0.0702 & 0.4547 & 0.3544 &0.6919 & 0.5939 & 0.0285 &  0.3709 &  0.3079 & 0.6117 &  0.4959\\
1.0  &0.008 & 0.1588 & 0.5002 & 0.3680 & 0.7276 & 0.6558 & 0.1177 &  0.4144 &  0.3211 & 0.6453 & 0.5449\\
     & 0.02 & 0.2671 & 0.5921 & 0.4146 & 0.7956 &  0.7742 & 0.2310 &  0.5072 &  0.3674 &0.7141 & 0.6425\\
      & 0.05 & 0.3729 & 0.7098 & 0.4786  &0.8880 & 0.9811 & 0.3243 &  0.6104 &  0.4247 & 0.7966 & 0.8037\\
\hline
      &0.0004&$-$0.0343 & 0.6018 & 0.4644 & 0.8728 & 1.6527 & $-$0.0243 &  0.4899 &  0.3989 &  0.7656 & 1.3136\\
      &0.004 & 0.2133 & 0.7365 & 0.5135 &0.9884 &  2.1862 & 0.1495 &  0.6234 &  0.4517 & 0.8811 & 1.7116\\
5.0 &0.008 & 0.2960 &  0.7743 & 0.5315 & 1.0436 & 2.7217 & 0.2024 &  0.6582 &  0.4669 &  0.9256 &  2.0744\\
      & 0.02 & 0.4687 & 0.8681 & 0.5774 & 1.1130 & 3.3731 & 0.3356 &  0.7553 &  0.5146 & 0.9977 & 2.5746\\
      & 0.05 & 0.7418 & 0.9876 & 0.6578 & 1.2468 & 4.4811 & 0.5682 &  0.8775 &  0.5909 &1.1224 &  3.3844 \\
\hline
        &0.0004& 0.0083 & 0.6656 & 0.4957 & 0.9230 &  2.6917 & $-$0.0060 &  0.5738 &  0.4413 &  0.8337 &  2.1660\\
        &0.004  & 0.2813 & 0.8011 & 0.5562 & 1.0814 & 3.5464  & 0.1869 &  0.7050 &  0.5033 & 0.9863 & 2.8331 \\
10.0 &0.008 & 0.4050 & 0.8515  & 0.5820 & 1.1570 & 4.3613  & 0.2790 &  0.7566 &  0.5288 & 1.0558 &  3.4528\\
      & 0.02 & 0.6073 & 0.9454 & 0.6228 &1.2018 &  5.7147  & 0.4574 &  0.8543 &  0.5714 &1.1047 &  4.5192\\
       &0.05  & 0.8814 & 1.0505 & 0.6939 & 1.3174 & 7.9907 & 0.7263  &  0.9669  &  0.6404  &1.2137 & 6.2489 \\
\hline
\hline
Age & $Z$& $(U-B)$ & $(B-V)$ & $(V-R)$ & $(V-I)$ & $\log_{10}(M/L_V)$ & $(U-B)$ & $(B-V)$ & $(V-R)$ &$(V-I)$& $\log_{10}(M/L_V)$ \\
(Gyr)& & \multicolumn{5}{c}{BC03 models with the Canonical IMF} &\multicolumn{5}{c}{Our models with the Canonical IMF}\\
\hline
     &0.0004&$-$0.1419 & 0.1914 & 0.2484 & 0.4542 & 0.2079 & $-$0.1990 &  0.1358 &  0.2048 &  0.3805 & 0.1715\\
     & 0.004 &$-$0.0073 & 0.3060 & 0.2825 & 0.5804 & 0.2140 & $-$0.0862 &  0.2266 &  0.2330 &  0.4894 & 0.1731\\ 
0.5 &0.008& 0.0782 & 0.3219 & 0.2860 & 0.5877 & 0.2351 & 0.0020 &  0.2451 &  0.2376 &  0.4984 & 0.1887\\
     & 0.02  & 0.1958  &0.3463 & 0.2962 & 0.5994 & 0.2688 & 0.1043 &  0.2701 &  0.2481 &  0.5128 & 0.2128\\
      & 0.05 & 0.2780 & 0.4425 & 0.3504 & 0.6855 & 0.3129 & 0.1912  & 0.3487  & 0.2951  & 0.5867 &  0.2494 \\
\hline
      &0.0004&$-$0.1389 &  0.3558 & 0.3301 & 0.6081 & 0.3048 & $-$0.1649 &  0.2764 &  0.2802 &  0.5255 & 0.2540\\
      & 0.004 &0.0693 & 0.4524 & 0.3505 & 0.6842 & 0.3350  & 0.0306 &  0.3702 &  0.3050 &  0.6059 & 0.2800\\
1.0  &0.008& 0.1586 &  0.4989 & 0.3654 & 0.7213 & 0.3703  & 0.1205 &  0.4137 &  0.3185 &  0.6392 & 0.3072\\
     & 0.02 & 0.2678 & 0.5921 & 0.4135 & 0.7912 & 0.4379 & 0.2321 &  0.5074 &  0.3667 &  0.7105 & 0.3640 \\
      & 0.05 &0.3736 & 0.7102 & 0.4780 & 0.8846 & 0.5530 & 0.3262 &  0.6133  & 0.4253 &  0.7954 & 0.4544 \\
\hline
      &0.0004&$-$0.0410 &  0.5917 &  0.4526 & 0.8544 &  0.9219 & $-$0.0295 &  0.4824 &  0.3885 &  0.7487 & 0.7307\\
      &0.004 &  0.2075 & 0.7291 & 0.5022 & 0.9684  & 1.2150  & 0.1494 &  0.6198 &  0.4427 &  0.8641 &  0.9518 \\
5.0 &0.008& 0.2902 & 0.7677 & 0.5222 & 1.0236 & 1.5087 & 0.1978 &  0.6529 &  0.4588 &  0.9073 & 1.1448\\
     & 0.02 & 0.4650 & 0.8645 & 0.5714 & 1.0953 & 1.8650 & 0.3331 &  0.7527 &  0.5094 &  0.9818 & 1.4190 \\
      & 0.05 &0.7392 & 0.9856 & 0.6534 & 1.2326 & 2.4681 & 0.5670  & 0.8760 &  0.5871 &  1.1098 & 1.8605 \\
\hline
      &0.0004& $-$0.0029 & 0.6511 & 0.4781 & 0.8958 & 1.5667 & $-$0.0149 &  0.5645 &  0.4273 &  0.8112 & 1.2618\\
      &0.004 &0.2733 & 0.7913 & 0.5404 & 1.0553 & 2.0177 & 0.1882 &  0.7013 &  0.4911 &  0.9649 & 1.6143\\
10.0 &0.008 &0.3989 & 0.8441 & 0.5700 & 1.1329 & 2.4420 & 0.2782 &  0.7534 &  0.5200 &  1.0372 & 1.9426\\
      & 0.02 & 0.6034 & 0.9411 & 0.6143 & 1.1759 & 3.1681 & 0.4581 &  0.8528 &  0.5652 &  1.0837 & 2.5103 \\
       &0.05  & 0.8793 &  1.0486 & 0.6872 & 1.2951 & 4.4130 & 0.7305 &  0.9684 &  0.6365 &  1.1975  & 3.4755 \\
\hline
\hline
\end{tabular}
\end{table*}

\section{Interpretation of cluster colours}

Although it is impossible for our models to provide the real BS
distribution in each individual star cluster, if our models are
reasonable, they should be able to reproduce the general trend of the
BS-induced modifications to the integrated-light properties of star
clusters. Thus, as the fundamental test of our models, we compare our
model results with the empirical conclusion from our previous
work. Our main conclusion was that the BS-induced colour changes are
roughly 0.10$\pm$0.12~mag in $(U-B)$, 0.09$\pm$0.09~mag in $(B-V)$ and
0.05$\pm$0.05~mag in $(V-R)$, based on a working sample of Galactic
OCs (Xin et al. 2007) and the Large Magellanic Cloud (LMC) star cluster
ESO~121--SC03 (Xin et al. 2008, their table 1).

\begin{figure*}
\includegraphics[width=6.5cm]{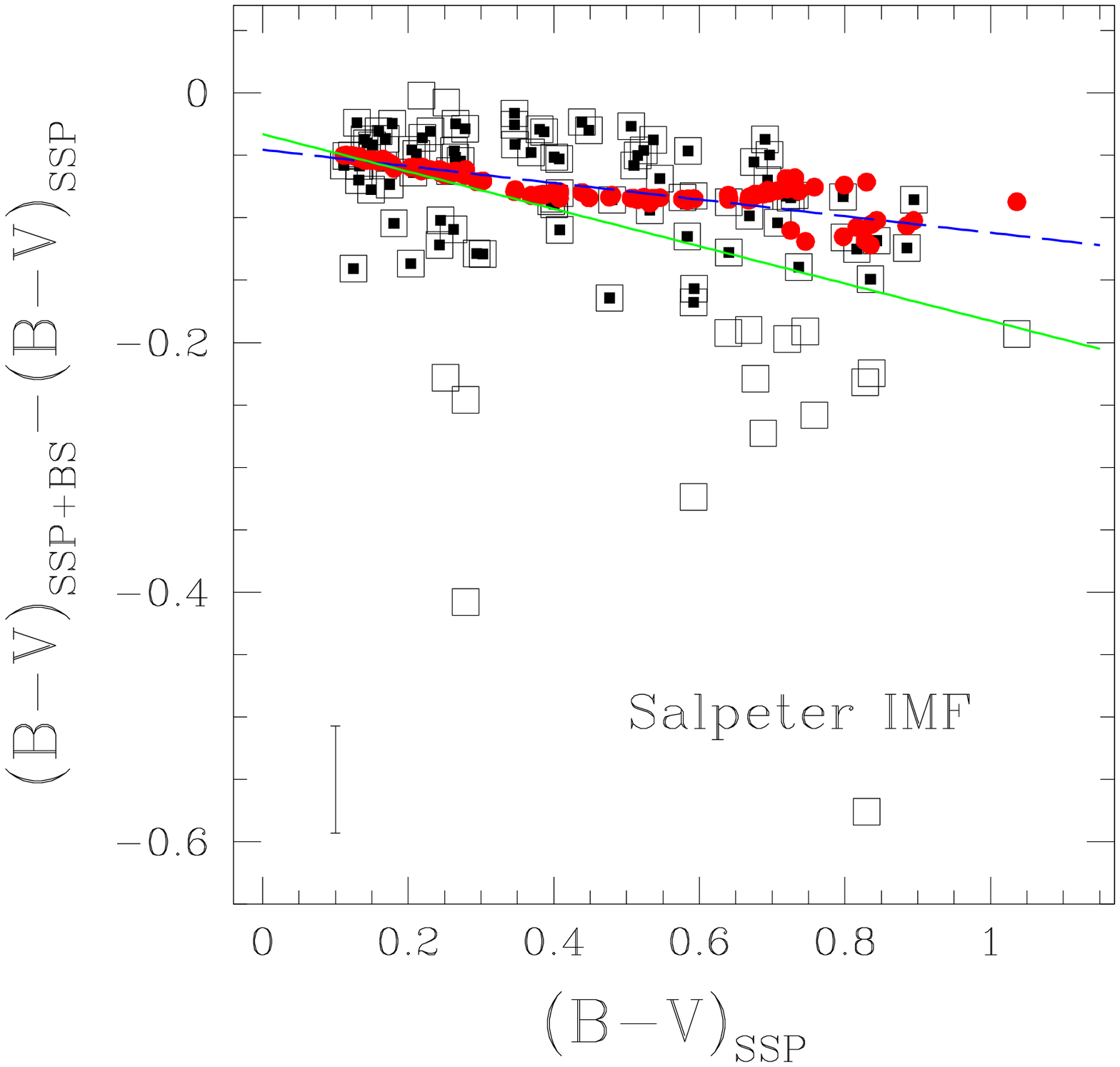}\includegraphics[width=6.5cm]{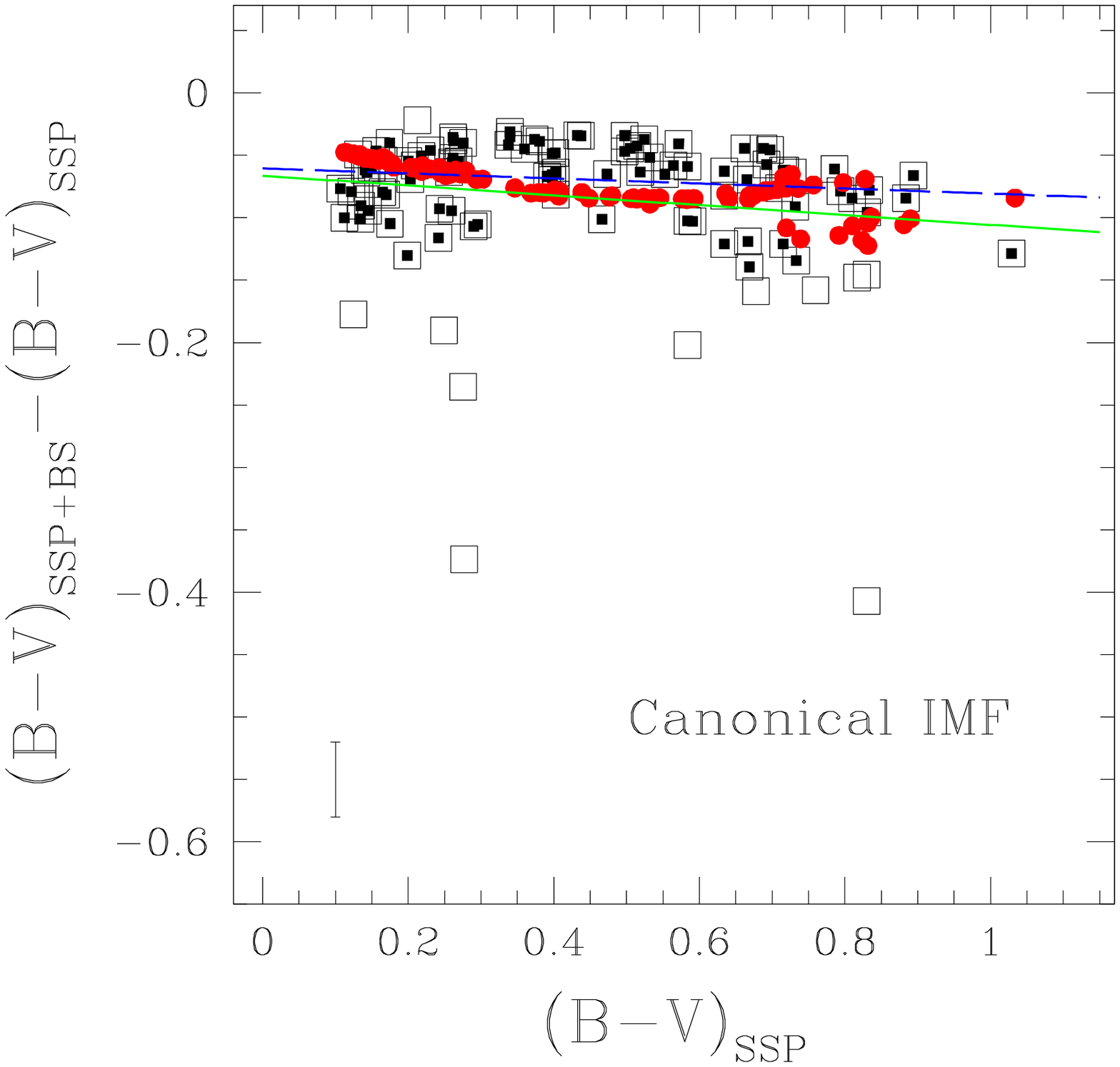}\\
\includegraphics[width=6.5cm]{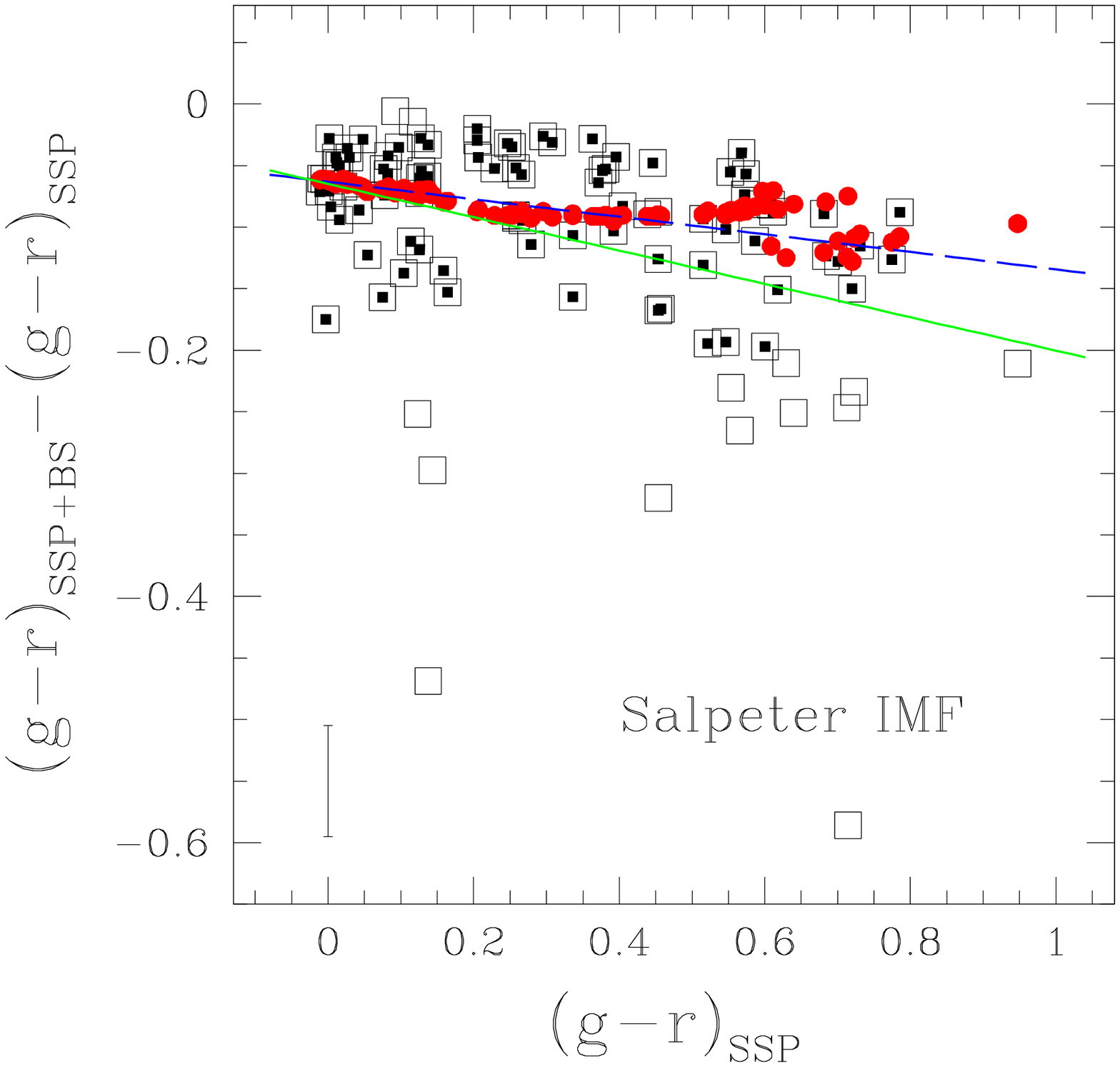}\includegraphics[width=6.5cm]{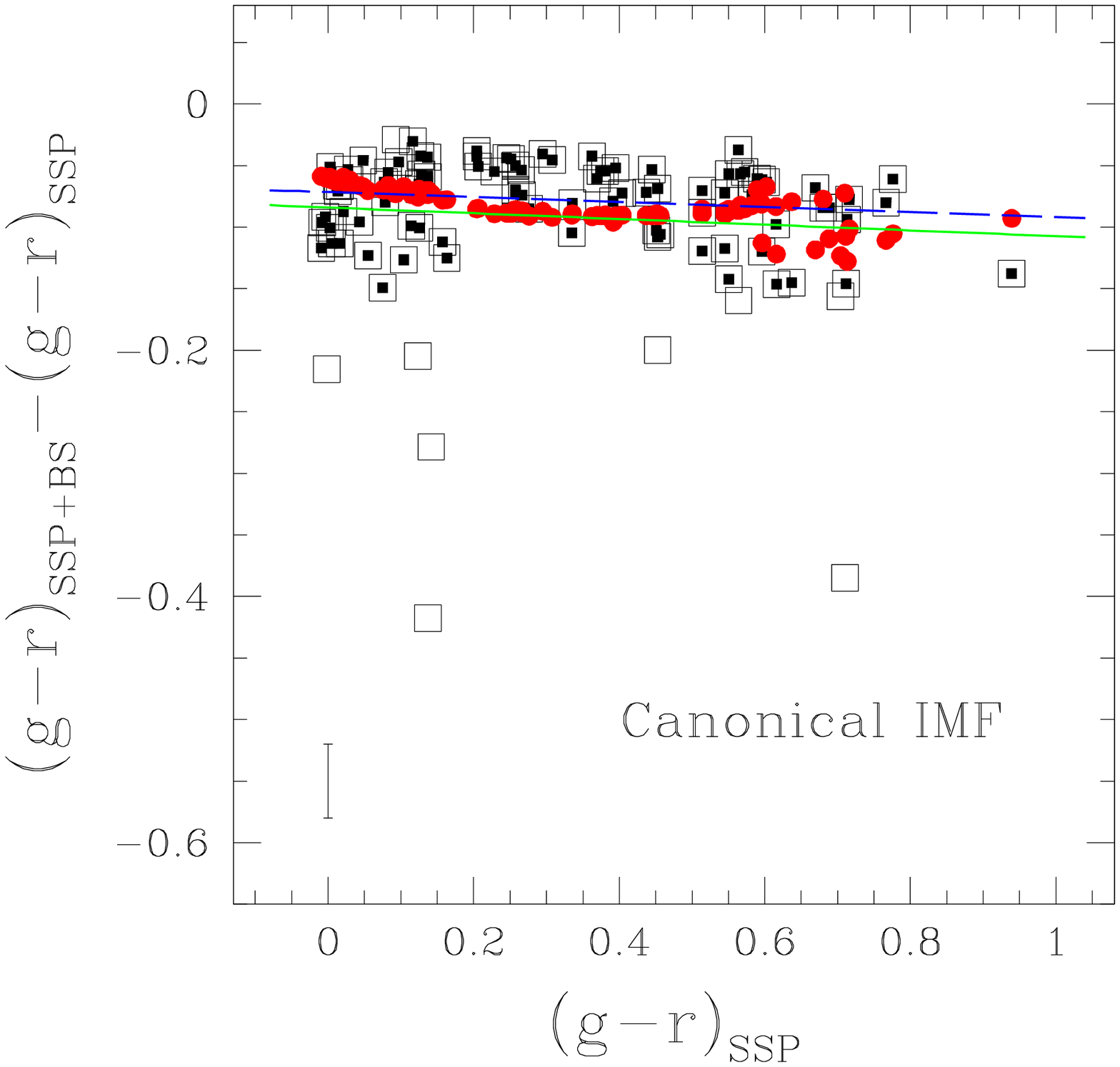}\\
\caption{Test of the consistency between our models and the
  statistical results from individual OCs. 
 The $x$ axis shows colours of the traditional SSPs. 
The $y$ axis shows BS-induced colour modifications. 
 The open squares represent the observational colour modifications of the 
100 Galactic OCs in our sample. The filled squares mark OCs with 
 colour variations less than the average plus $1\sigma$. 
The dashed line in each panel is the least-squares fit to the filled squares, 
and the solid line is the least-squares fit to all the open squares. 
The filled circles represent the corresponding results from SSP models.}
\label{fig12}
\end{figure*}

Fig.~\ref{fig12} is designed to test the consistency between our
models and the empirical results from individual OCs.  The $x$ axis
shows colours of the traditional SSPs, the $y$ axis shows BS-induced
colour modifications, and the open squares represent the observational
colour modifications of the 100 Galactic OCs in our sample. We
calculated the standard deviation ($\sigma$) of the colour variation
for the 100 OCs, e.g., $\sigma_{(B-V)}=\sqrt
{\frac{\sum_{i=1}^N(\Delta(B-V)_i-\overline{\Delta(B-V)})^2}{N\times(N-1)}}$
and $N=100$. The filled squares mark OCs with colour variations within
$1\sigma$, i.e., $abs(\Delta(B-V)) \le
\overline{\Delta(B-V)}+1\sigma$.  In all panels, the error bars show
the $1\sigma$ values and the dashed lines are the least-squares fits
to the filled squares. The solid lines are the least-squares fits to
all open squares. The filled circles are the corresponding results
from SSP models, i.e., the $y$-axis value of the filled circles
represents the differences between our models and BC03.  The
broad-band Johnson-Cousins $(B-V)$ (top panels) and Sloan Digital Sky
Survey (SDSS) $(g-r)$ colours (bottom panels) for both IMFs are
included in the figure. The colour differences for different IMFs are
caused by differences in the normalisation constants adopted to
retrieve the real intensity of a given SSP. These normalisation
constants are obtained using the observed $N_2$ values of OCs, as
defined by Eq.~(\ref{eq4}). We find good agreement between our model
predictions and the empirical results for Galactic OCs.  Fluctuations
dramatically decrease in our model results compared to the
observations.  The small dispersion in $\Delta(B-V)$ and $\Delta(g-r)$
starting at $(B-V) \sim 0.7$ and $(g-r) \sim 0.6$ mag in our model
results (solid circles) may be intrinsic, i.e., for star clusters
older than $\sim 5.0$ Gyr -- roughly $(B-V) \sim 0.7$ mag --
broad-band colours are more sensitive to metallicity than to age
(Harris et al. 2006), and a metallicity dispersion is observed among
our sample OCs.

\section{Interpretation of cluster spectra}

\subsection{UV spectra}

\begin{figure}
\begin{center}
\includegraphics[width=8.5cm]{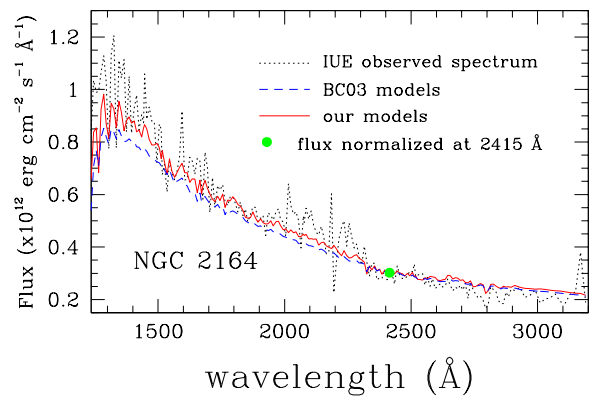} 
\caption{Understanding the IUE spectrum of NGC 2164 based on both BC03
and our models.  The dotted line is the observed spectrum, the solid
line is the spectrum from our models and the dashed line is the BC03
spectrum.}
\label{fig13}
\end{center}
\end{figure}

Given the typical loci of BSs in cluster CMDs, our models may be
expected to be more useful in the interpretation of observed UV
spectra of star clusters. Fig.~\ref{fig13} shows a test using the
observed UV spectrum of the LMC star cluster NGC~2164. The spectrum is
extracted from the International Ultraviolet Explorer (IUE)
archive. The intrinsic flux of the spectrum is obtained using the
extinction law described in Cassatella et al. (1987). The fundamental
parameters of the cluster adopted for the comparison are an age of 0.2
Gyr and $Z=0.004$ (Vallenari et al. 1991). The dotted line in the
figure is the observed spectrum. The solid line is the spectrum from
our models based on the cluster's parameters.  The dashed line is the
corresponding BC03 spectrum.

The flux of model spectrum is normalised to the observed value at 2415
{\AA} (filled circle). The normalisation position is chosen because
the flux fluctuations are relatively small at this wavelength. There
is no dramatic difference between the spectra of the two SSP
models. They are nearly identical at $\sim$ 3150--2500~{\AA}. Our
model spectrum is brighter at wavelengths shortward of 2415 {\AA} and
can fit the observation better compared to BC03 at wavelengths of
$\sim$ 2000--1700 {\AA}; however, our model is not sufficiently bright
to fit the observed spectrum at wavelengths shorter than 1500 {\AA}
(although still better than BC03).

\subsection{Optical spectra}

\begin{figure}
\begin{center}
\includegraphics[width=8.5cm]{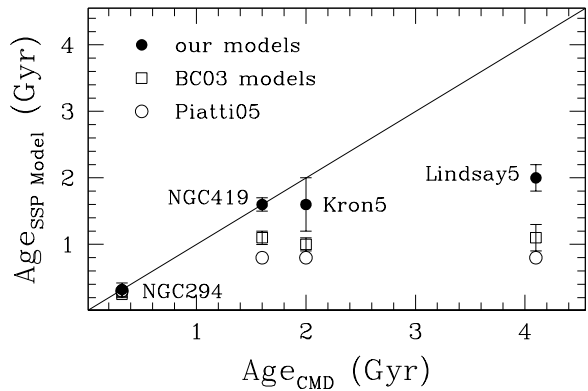}
\caption{Comparison of the age determinations from ($x$ axis)
  isochrone fits and ($y$ axis) spectral fits using different SSP
  models. The corresponding values are listed in Table~\ref{table4}.}
\label{fig14}
\end{center}
\end{figure}

Piatti et al. (2005a) published flux-calibrated integrated spectra of
18 Small Magellanic Cloud (SMC) star clusters covering the wavelength
range of 3600--6800 {\AA}.  This provides an ideal data set to test
our models.  The observed integrated spectra of Magellanic Cloud star
clusters are better sampled, i.e., the observed spectra more
accurately represent the clusters' real spectra, than the spectra of
Galactic star clusters, simply because of their greater distances
(which facilitates straightforward observations of the entire objects
in a single observation). For inclusion in our sample, a star cluster
needs to have an age and metallicity within the respective ranges
covered by our models. It also needs to have a reliable age
determination from CMD analysis. Based on these two requirements, nine
of the 18 star clusters (Piatti et al. 2005a, their table 3) meet our
selection criteria for the model test, including Lindsay~5, Kron~3, 5,
6, 7 and 28, and NGC~411, NGC~419 and NGC~458. The observed spectra of
all nine clusters are available in Piatti et al (2005a).

Since there is no specific extinction law mentioned in their paper,
we use the normal one, i.e., $A_{\lambda}=0.65\times
A_V(1/\lambda-0.35)$ and $A_V=3.1\times E(B-V)$ (Bica \& Alloin 1986),
to correct the observed spectra for reddening and, since our model
spectra are constructed using theoretical stellar spectra at a low
resolution of 15~{\AA}, the observed spectra are degraded with a bin
size of 30~{\AA}. Our model spectra are normalised to the same
wavelength grid as the degraded observed spectra. We next identify the
continuum of the latter. Since the BS spectra in our models are
approximated by spectra of single MS stars, our models will not be
able to fully account for all changes in the SSP spectra caused by
BSs, such as detailed modifications to the spectral lines that are
related to BS formation scenarios. Therefore, we use the continuum
instead of the entire spectra for the model test.

\begin{table*}
\caption{Comparison of age measurements of four SMC star clusters}
\label{table4}
\begin{tabular}{lllcllcc}
\hline
\hline
Cluster name& Age$_{\rm CMD}$ & Ref. &Age$_{\rm Piatti05}$ &Age$_{\rm BC03}$ & Age$_{\rm our\; model}$ & [Fe/H]& Ref. \\ 
                                                     &(Gyr) &&                   (Gyr) & (Gyr) & (Gyr)&&\\
\hline
Kron~5 & 2.0& 1 & 0.8&1.0$\pm$0.1&1.6$\pm$0.4&$-$0.60&1\\
Lindsay~5 &  4.1 & 1 &0.8&1.1$\pm$0.2&2.0$\pm$0.2&$-$1.20&1\\
NGC~294 &0.32$\pm$0.15&2&0.3&0.25$\pm$0.05&0.32$\pm$0.10&$-$0.95&2\\
NGC~419 &2.0$\pm$0.2, 1.6$\pm$0.4 &3, 4&0.8&1.1$\pm$0.1&1.6$\pm$0.1&$-$0.70&4\\
\hline
\end{tabular}\\
References: 1. Piatti et al. (2005b); 2. Piatti et al. 2007; 3. Rich
et al. (2000); 4. Piatti et al. (2002).
\end{table*}

\begin{figure*}
\begin{center}
\includegraphics[width=18cm]{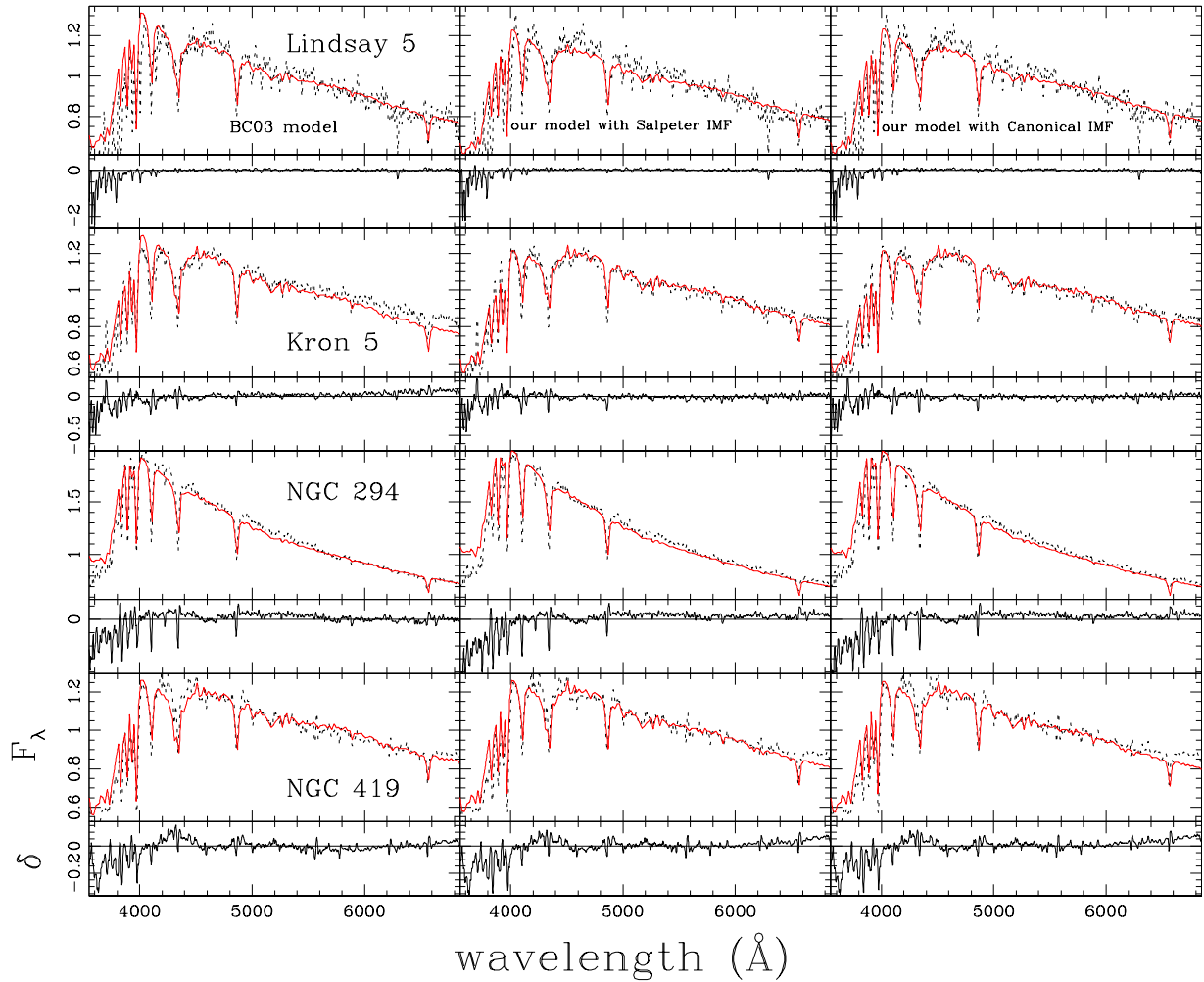}
\caption{Fit to the observed spectra of SMC star clusters based on
  different SSP models. The dotted lines are the observed, integrated
  spectra. The solid lines are the spectra from different SSP
  models. The residuals of the fit, $\delta=(F(\lambda)_{\rm
    observed}- F(\lambda)_{\rm model})/F(\lambda)_{\rm observed}$, are
  given in the bottom part of each panel. }
\label{fig15}
\end{center}
\end{figure*}

Contrary to our UV spectra test discussed in previous subsection,
we leave the cluster age as a free parameter in this test. We fit the
observed spectra with the model spectra for different ages and
calculate the standard deviation for selected points of the continuum
between the observed and model spectra to identify the best fit. The
best fit yields a predicted age for the cluster. We subsequently
compare the age values from BC03 and from our models to see if our
models can derive more accurate ages for our sample clusters, i.e.,
ages that are closer to the age based on CMD isochrone fitting.

As a result, our models give more accurate age determinations for four
SMC star clusters from among the nine sample objects, Lindsay~5,
Kron~5, NGC~294 and NGC~419.  As for the other five SMC star clusters
(i.e., Kron~3, 6, 7 and 28, and NGC~458), their observed spectra
cannot be matched by either the BC03 or our models. Kron~3, 6, 7 and
28 are characterised by very high intensities at the red ends of their
spectra, which can only be traced by our models if we adopt ages that
are much older ($> 10$ Gyr) than the clusters' real ages (from CMD
analysis). The BC03 models cannot fit their spectra in the blue and
red bands simultaneously. Explanations such as the absence of
AGB stars from both models, the presence of carbon stars (Feast \&
Lloyd~Evans 1973; Mould et al. 1992), and contamination by field stars
(Piatti et al. 2001) may be responsible for the poor understanding of
the observed spectra of these five clusters. Since this is beyond the
scope of this paper, we focus instead on the four clusters which can
be readily understood on the basis of SSP model analysis.

The detailed fit results of these four clusters are included in
Table~\ref{table4}, where columns (2)--(3) give the clusters' CMD ages
and the corresponding references, respectively, column (4) lists the
ages from Piatti et al. (2005a) based on their own empirical SSP
templates using observed spectra of Galactic OCs and GCs (Bica \&
Alloin 1986), columns (5)--(6) present the ages from BC03 and from our
models, respectively, and the adopted [Fe/H] value for the fit to the
spectrum and the corresponding references are given in columns
(7)--(8), respectively. A direct comparison of the age determinations
for the four SMC star clusters using different SSP models is presented
in Fig.~\ref{fig14}.

Fig.~\ref{fig15} shows details of the spectral fits to the four SMC
star clusters. The dotted lines are the observed spectra. The solid
lines are the best-fitting spectra from different SSP models. The fit
residuals are presented as the solid lines in the bottom part of each
panel. They have been calculated as $\delta=(F(\lambda)_{\rm
observed}- F(\lambda)_{\rm model})/F(\lambda)_{\rm observed}$. Both
the BC03 models and our models with different IMFs can perfectly fit
the continuum of the four SMC star clusters over almost the entire
wavelength range. The model spectra can also roughly match the
features of the Balmer lines of the observed spectra.  The best fit
for different models refers to different age determinations for the
same star cluster. BC03 always returns a younger age for a given
cluster compared to the age from our models (see Table~\ref{table4}),
since BSs turn the SSPs hotter.  Considerable uncertainties only
appear for $\lambda < 4000~{\AA}$, where flux fluctuations become
larger because the number of photons dramatically drops and the CCDs
used for such observations usually have low quantum efficiency at UV
wavelengths. Meanwhile, SSP models also have problems in correctly
presenting ISEDs in the UV because of limitations such as those
associated with the adopted stellar-atmosphere models, which may also
be responsible for the poor fit at UV wavelengths.

Using the technique of isochrone fitting is comparably easier to
identify the dominant population of a star cluster and remove
contamination by field stars, and thus to obtain a better estimate of
the age and/or metallicity of the cluster. If the parameters are
predicted by comparing the observed spectra with model spectra, the
results will always suffer from field contamination, and possibly from
the presence of multiple populations in the cluster, as well as from
limitations inherent to the model coverage. Therefore, our conclusion
from the model test is that the BS-SSP models can yield more accurate
age values for stellar population studies in {\it unresolved
  conditions}, compared to the ages derived from conventional SSP
models.

We did not attempt a metallicity derivation in our model test, mainly
because metallicity information is more commonly derived from the
properties of a range of characteristic spectral lines, and the
current BS-SSP models are not suitable for this purpose. However, if
one wants to estimate the metallicity to first order by fitting the
continuum of the observed spectrum with model spectra of different
metallicities (more metal-poor SSPs have stronger, higher-intensity
ISEDs; see the bottom right-hand panel of Fig.~\ref{fig9}), BSs would
exhibit the same effect as seen for the age determination, i.e., BSs
make the ISED hotter compared to ISEDs from conventional SSPs models
for a given metallicity, which means that the apparent metal-poor
spectrum of a star cluster could actually be more metal-rich.

\section{Summary and discussion}

In this paper, we have presented a new set of SSP models that include
the contributions of blue straggler stars, i.e., `BS-corrected' SSP
(BS-SSP) models. The new models are developed to improve the standard
SSP models by taking into account the effects of stellar interactions
through empirical inclusion of the BS contribution in each SSP. The
empirical properties of the BS populations are obtained based on
photometric data of BSs in 100 Galactic OCs.

We have tested that (i) the broad-band colours in our models are
highly consistent with the colours of 100 Galactic OCs if BSs are
properly accounted for and (ii) age predictions based on spectral fits
to our models yield ages that are closer to the CMD ages compared to
the ages derived based on the BC03 models. The test results prove the
reliability of our models and indicate that our models offer important
improvements for studies of stellar populations.

The basic description of our BS-SSP models includes:

\begin{enumerate}

\item The models cover the wavelength range from 91{\AA} to 160
  $\mu$m, ages from 0.1 to 20 Gyr and metallicities $Z=0.0004, 0.004,
  0.008, 0.02$ (solar metallicity) and 0.05. The metallicity
  $Z=0.0001$ is not included, because extended HB 
  stars, instead of BSs, dominate the energies in the UV and blue
  bands in such extremely metal-poor SSPs.

\item The models are constructed as increments to the BC03 standard
  SSP models using the Padova1994 isochrones and the Lejeune et
  al. (1997) stellar spectra,. They can thus be used directly in EPS
  studies as replacement of BC03 for the same parameter
  coverage. Application of the models should be limited to the
  `low-resolution' regime. As each BS spectrum is approximated by the
  theoretical spectrum of a single MS star, the models cannot fully
  account for changes in the spectral lines that are related to the
  formation scenarios of BSs.

\item The essential effect of BSs is to make an SSP's ISED hotter in
  the UV, blue and optical bands, and consequently turn the broad-band
  colours much bluer. Taking the SSP models with $Z=0.02$ as an
  example, the differences in the broad-band colours between BC03 and
  our models are 0.10$\pm$0.05~mag in $(U-B)$, 0.08$\pm$0.02~mag in
  $(B-V)$, 0.05$\pm$0.01~mag in $(V-R)$, 0.12$\pm$0.05~mag in $(u-g)$,
  0.09$\pm$0.02~mag in $(g-r)$ and 0.17$\pm$0.03~mag in $(g-z)$.
\end{enumerate}

Given the common presence of BSs in various stellar systems, we
believe that the BS-SSP models will enable the community to uncover
interesting results in studies of stellar populations. In general, our
models allow old SSPs to have higher energies (intensities) in UV and
blue bands, which implies that the prediction from our models will be
in the direction of older age and/or higher metallicity for a given
population compared to the standard SSP models. For instance, (i) the
shift of $\sim0.10$ mag towards the blue for the broad-band colours
may influence our understanding of the age of the oldest stellar
population in a galaxy based on the observed colour distribution of
GCs (e.g., Yoon et al. 2006, who quote a typical observational error
of $\sim0.10$ mag in $(g-z)$); and (ii) the dramatic enhancement in
the ISEDs caused by BSs may lead to a new understanding of the UV
upturn in elliptical galaxies. Han et al. (2007) successfully
approached the problem with their newly developed EPS models that now
also include binary evolution. Since binary interaction is also one of
the major formation mechanisms of BSs, it is possible that BSs can (at
least partially) contribute to this UV excess.

The current version of the BS-SSP models is available from
http://www.astro.uni-bonn.de/{\textasciitilde}webaiub/english/downloads.php
and http://sss.bao.ac.cn/bss. We will update the models when any
significant and systematic improvements based on either observational
or theoretical studies regarding BS formation in star clusters become
available. Updated versions of the models will be provided at
http://sss.bao.ac.cn/bss.

\section*{Acknowledgments}

YX gratefully acknowledges financial support from the Alexander von
Humboldt Foundation. YX also thanks the Chinese National Science
Foundation (NSFC) for support through grant 10773015. 
LD thanks NSFC grant 10973015, and the Ministry of Science 
and Technology of China grant 2007CB815406. 
RdG acknowledges NSFC grant 11043006.

\bsp


\end{document}